\documentclass[twocolumn,english,xlinenumbers,aps,twocolumn]{revtex4-1}
\usepackage[T1]{fontenc}
\usepackage[utf8]{inputenc}
\usepackage{color}
\usepackage{babel}
\usepackage{amsmath}
\usepackage{amssymb}
\usepackage{graphicx}
\usepackage[unicode=true,pdfusetitle,
 bookmarks=true,bookmarksnumbered=false,bookmarksopen=false,
 breaklinks=false,pdfborder={0 0 0},pdfborderstyle={},backref=false,colorlinks=true]
 {hyperref}
\hypersetup{
 pdfborderstyle={},citecolor=blue}

\makeatletter

\providecommand{\tabularnewline}{\\}

\definecolor{darkblue}{rgb}{0.0,0.0,0.5}

\graphicspath{{./}{./FIGURES/}} 

\makeatother

\begin{document}

\title{Dynamics and instabilities of Lasing Light Bullets in Passively Mode-Locked Semiconductor Lasers}

\author{S. V. Gurevich}
\email{gurevics@uni-muenster.de}

\altaffiliation{Center for Nonlinear Science (CeNoS), University of Münster, Corrensstrasse 2, D-48149 Münster, Germany}
\affiliation{Institute for Theoretical Physics, University of Münster, Wilhelm-Klemm-Str. 9, D-48149 Münster, Germany}

\author{J. Javaloyes}

\affiliation{Departament de Física, Universitat de les Illes Balears, C/ Valldemossa
km 7.5, 07122 Mallorca, Spain}
\begin{abstract}
Recently, the existence of robust three-dimensional light bullets
(LBs) was predicted theoretically in the output of a laser coupled
to a distant saturable absorber. In this manuscript, we analyze the
stability and the range of existence of these dissipative localized
structures and provide guidelines and realistic parameter sets for
their experimental observation. In order to reduce the complexity
of the analysis, we first approximate the three-dimensional problem
by a reduced equation governing the dynamics of the transverse profile.
This effective theory provides an intuitive picture of the LB formation
mechanism. Moreover, it allows us to perform a detailed multi-parameter
bifurcation study and to identify the different mechanisms of instability.
It is found that the LBs experience dominantly either homogeneous
oscillation or symmetry breaking transversal waves radiation. In addition,
our analysis reveals several non-intuitive scaling behaviors as functions
of the linewidth enhancement factors and the saturation parameters.
Our results are confirmed by direct numerical simulations of the full
system.
\end{abstract}
\maketitle

\section{Introduction}

Light bullets (LBs) consists in pulses of light that are simultaneously
confined in the transverse and the propagation directions. In the
context of dissipative systems, LBs can be considered as Localized States
(LSs) and are thus attractors of the dynamics. These hypothetical
objects attracted a lot of interest in the last twenty years, both
for fundamental and practical reasons. In practice, LBs should be
addressable, i.e., they could be individually turned on and off, and
one can envision that they would circulate indefinitely within an
optical cavity as elementary bits of information.

Traditionally, the optical confinement scenario that would lead to
LBs is envisioned trough a conservative mechanisms in which a self-focusing
nonlinearity compensates for the spreading effect of chromatic dispersion
and/or diffraction. Seminal works demonstrated however that if the
number of spatial dimensions is too large $\left(d_{\perp}\ge2\right)$,
LBs are unstable and collapse \cite{S-OL-90}, which was a result
earlier discovered in the field of plasma physics \cite{Z-SJETP-72}.
Other confinement mechanisms were envisioned in forced dissipative
system and LBs were predicted in optical parametric oscillators \cite{TM-PRA-99}
and bistable cavities \cite{VVK-OS-00,BMP-PRL-04,CPM-NJP-06} with
instantaneous nonlinearities.

Recently, a regime of temporal localization was predicted and experimentally
demonstrated in a semiconductor passively mode-locked laser \cite{MJB-PRL-14}.
Passive mode-locking (PML) is a well known method for achieving short
optical pulses \cite{haus00rev}. It is achieved by combining two
elements, a laser amplifier providing gain and a nonlinear loss element,
usually a saturable absorber. For proper parameters, this combination
leads to the emission of temporal pulses much shorter than the cavity
round-trip $\tau$. It was shown in \cite{MJB-PRL-14} that, if operated
in the long cavity regime, the PML pulses become individually addressable
temporal LSs coexisting with the off solution. In this long cavity
regime, the round-trip time $\tau$ is made much longer than the semiconductor
gain recovery time $\tau_{g}\sim1\,$ns, which is slowest variable.
Interestingly, this temporal localization regime was found to be compatible
with an additional spatial confinement mechanism, which lead to the
theoretical prediction of a regime of stable three dimensional LBs
\cite{J-PRL-16}.

While preliminary results based upon direct numerical integration
allowed finding some basic estimates of the stability range for a
generic parameter set, a full bifurcation study of the system described
in \cite{J-PRL-16} is still lacking. Yet, a multi-parameter analysis
considering the various design factors of a passively mode-locked
laser system would be of high relevance, in particular to experimental
groups, as it would inform upon the proper parameter ranges in which
an experimental realization may take place. In particular, assessing
not only the range of existence of the LBs but also their destabilization
mechanisms is of paramount importance. However, the LBs presented
in \cite{J-PRL-16} are particularly stiff multiple timescale objects
in which the optical pulse is followed by a material ``trail'' that
differs in extension by three orders of magnitudes. This stiffness,
that occurs in the temporal domain or equivalently along the propagation
axis, is exacerbated by the presence of the transverse dimensions
that make a bifurcation analysis of two and three dimensional LBs
a challenging problem.

We perform this analysis in this manuscript in two steps. Firstly,
we approximate the solutions of the 3D problem by the product of a
slowly evolving transverse profile and of a short pulse propagating
inside the cavity. This allows us to obtain a reduced model governing
the dynamics of the transverse profile. This effective theory allows
to consider the LBs as if they were static diffractive spatial auto-solitons,
similar, e.g., to those in \cite{RK-OS-88}. We show that the transverse
profile is governed by an effective Rosanov equation which allows
for a detailed multi-parameter bifurcation study and also to identify
the different mechanisms of instability. For that purpose, we employ
the continuation and bifurcation package pde2path \cite{pde2path}.
It is found that the the light bullets experience dominantly either
homogeneous oscillation or symmetry breaking lateral waves radiation.
In addition, our analysis reveals several non-intuitive scaling behaviors
as a functions of the linewidth enhancement factors and the saturation
parameter. In the second stage, our predictions are confirmed by extensive
direct numerical simulations of the spatio-temporal dynamics of the
full system.

\section{Model }

\begin{figure}[ht!]
\includegraphics[width=1\linewidth]{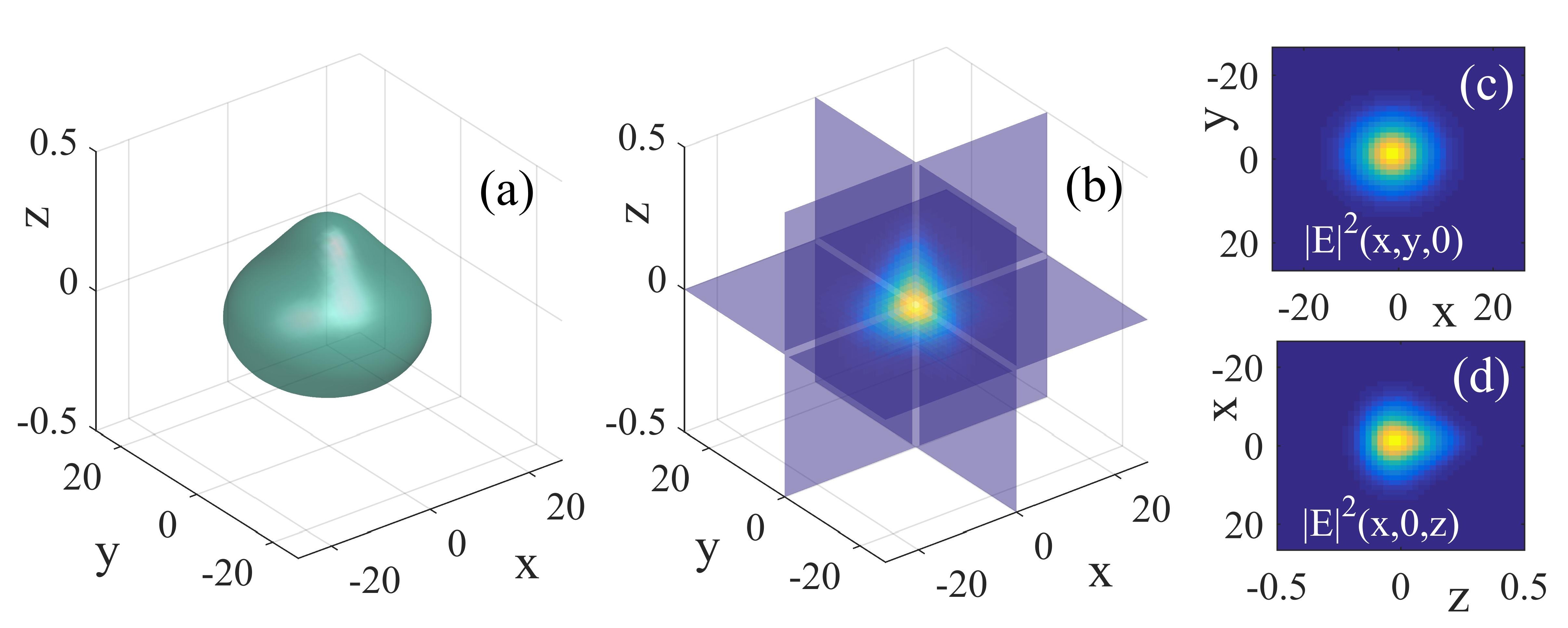}\caption{(Color online) Exemplary solutions of Eqs.~(\ref{eq:VTJ1}-\ref{eq:VTJ3})
showing the intensity profile of a stable LB. (a) Isosurface at $1\%$
of the maximal value. (b) Cross-sections in the three orthogonal planes
defined as $x=0$, $y=0$ and $z=0$. The corresponding cross-section
profiles are presented in the insets (c) and (d). Parameters are $\left(\gamma,\kappa,\alpha,\beta,\Gamma,G_{0},Q_{0},s,d\right)=\left(40,0.8,1.5,0.5,0.04,0.425,0.354,30,10^{-2}\right)$.}
\label{fig:LB3D} 
\end{figure}

We describe the passively mode-locked laser using the generic Haus
partial differential equation (PDE) \cite{haus00rev}. We consider
a situation in which a broad area gain chip is coupled to a distant
saturable absorber with telescopic optics in self-imaging conditions,
as for instance in \cite{GBG-PRL-08}. In this situation, each point
of the gain section is mapped onto the absorber section and vice-versa.
The diffraction in our system is the result from the propagation within
the active sections. We assume that it is assumed sufficiently small
as to justify the use of the paraxial approximation, as e.g. in \cite{BLP-PRL-97}.
We also work in the limit of moderate gain ($G$) and saturable absorption
($Q$) such that the uniform field limit applies. In this context,
the equation governing the evolution of the field profile $E\left(r_{\perp},z,\sigma\right)$
over the slow time scale $\sigma$ reads
\begin{eqnarray}
\partial_{\sigma}E & = & \left\{ \sqrt{\kappa}\left[1+\frac{1-i\alpha}{2}G\left(r_{\perp},z,\sigma\right)-\frac{1-i\beta}{2}Q\left(r_{\perp},z,\sigma\right)\right]\right. \nonumber \\
 & - & 1 + \left.\frac{1}{2\gamma^{2}}\partial_{z}^{2}+\left(d+i\right)\Delta_{\perp}\right\} E\left(r_{\perp},z,\sigma\right),\label{eq:VTJ1}
\end{eqnarray}
where $\gamma$ is the bandwidth of the spectral filter
representing, e.g., the resonance of a VCSEL \cite{MJB-JSTQE-15},
$\Delta_{\perp}=\partial_{x}^{2}+\partial_{y}^{2}$ is the transverse
Laplacian, $\kappa$ is the fraction of the power remaining in the
cavity after each round-trip and $\alpha$ and $\beta$ are the linewidth
enhancement factors of the gain and absorber sections, respectively.
The amount of diffraction in the combined gain and absorber sections
can be described by a diffraction length that was used in Eq.(\ref{eq:VTJ1})
to normalize the transverse space variables $r_{\perp}=\left(x,y\right)$.
As such, the transverse domain size $L_{\perp}$ becomes a bifurcation
parameter. For small $L_{\perp}$, the system is governed by its transverse
boundary conditions and conversely, localized states may occur when
$L_{\perp}\gg1$. The parameter $d$ represents the small amount of
field diffusion incurred for instance by the dependence of the reflectivity
of the VCSEL distributed Bragg Reflectors upon the angle of incidence.
The longitudinal variable $\left(z\right)$ is identified as a fast
time variable and represents the evolution of the field within the
round-trip. The carrier dynamics read 
\begin{eqnarray}
\partial_{z}G & = & \Gamma G_{0}-G\left(\Gamma+\left|E\right|^{2}\right)+\mathcal{D}_{g}\Delta_{\perp}G,\label{eq:VTJ2}\\
\partial_{z}Q & = & Q_{0}-Q\left(1+s\left|E\right|^{2}\right)+\mathcal{D}_{q}\Delta_{\perp}Q,\label{eq:VTJ3}
\end{eqnarray}
with $G_{0}$ the pumping rate, $\Gamma=\tau_{g}^{-1}$ the gain recovery
rate, $Q_{0}$ is the value of the unsaturated losses, $s$ the ratio
of the saturation energy of the gain and of the SA sections and $\mathcal{D}_{g,q}$
the scaled diffusion coefficients. In general, the non-instantaneous
and causal response of the active medium represented by the variable
$G$ implies a lack of parity along $\left(z\right)$ for the LSs
generated by Eqs.~(\ref{eq:VTJ1}-\ref{eq:VTJ3}), see \cite{JCM-PRL-16}
for details. In Eqs.(\ref{eq:VTJ1}-\ref{eq:VTJ3}) the fast time
$\left(z\right)$ has been normalized to the SA recovery time that
we assume to be $\tau_{sa}=20\,$ps. Setting $\gamma=40$ and $\Gamma=0.04$,
corresponds to a full Width at Half Maximum of $250\,$GHz for the
gain bandwidth and a carrier recovery time $\tau_{g}=500\,$ps. Assuming
a diffraction length of $l_{\text{\ensuremath{\perp}}}=1\,\mu$m and
a domain size $L_{\perp}=190$ corresponds to a $190\,\mu$m broad
area device. The typical dimensions of the LB are $l_{LB}\sim10\,\mu$m
and $\tau_{LB}\sim4\,$ps. Finally, it was shown in \cite{J-PRL-16}
that carrier diffusion plays almost no role in the dynamics, so that
we set the diffusion coefficients $\mathcal{D}_{g,q}=0$. For proper
system parameters, Eqs.~(\ref{eq:VTJ1}-\ref{eq:VTJ3}) sustain the
existence of stable three-dimensional light bullets as depicted in
Fig.~\ref{fig:LB3D}. The details regarding the numerical method
used to solve Eqs.~(\ref{eq:VTJ1}-\ref{eq:VTJ3}) can be found in
Section 4 of the Appendix.

Exploiting the seminal work of New \cite{N-JQE-74} and the fact that
the LBs are composed of variables evolving over widely different timescales,
one can find an approximate model governing the shaping of the transverse
profile. We assume that the field reads $E\left(r_{\perp},z,\sigma\right)=A\left(r_{\perp},\sigma\right)p\left(z\right)$
with $p\left(z\right)$ a short normalized temporal pulse of length
$\tau_{p}$ that represents the temporal LS upon which the LB is built
and $A\left(r_{\text{\ensuremath{\perp}}},\sigma\right)$ is a slowly
evolving amplitude. Separating the temporal evolution into the fast
and slow parts corresponding to the pulse emission and the subsequent
gain recovery allows us to find the equation governing the dynamics
of $A$ as 
\begin{eqnarray}
\partial_{t}A & = & (d+i)\left(\partial_{u}^{2}+\partial_{v}^{2}\right)A+f\left(\left|A\right|^{2}\right)A.\label{eq:Rosa1}
\end{eqnarray}
Defining $h\left(P\right)=\left(1-e^{-P}\right)/P$, $P=|A|^{2}$
the function $f$ reads
\begin{eqnarray}
\negthickspace\negthickspace\negthickspace f\left(P\right) & = & \left(1-i\alpha\right)g\left(1+q\right)h\left(P\right)-\left(1-i\beta\right)qh\left(sP\right)-1,\label{eq:Rosa2}
\end{eqnarray}
see Section 1 of the Appendix for more
details. We defined in Eqs.~(\ref{eq:Rosa1},\ref{eq:Rosa2}) the
scaled spatial and temporal coordinates as $t=\left(1-\sqrt{\kappa}\right)\sigma$
and $\left(u,v\right)=\sqrt{1-\sqrt{\kappa}}\left(x,y\right).$ The
effective parameters that are the gain normalized to threshold $g$
and the normalized absorption $q$ as $g=G_{0}/G_{th}$ and $q=Q_{0}/\left(\frac{2}{\sqrt{\kappa}}-2\right)$.
We defined $G_{th}=\frac{2}{\sqrt{\kappa}}-2+Q_{0}$ as the threshold
gain value above which the off solution $\left(E,G,Q\right)=\left(0,G_{0},Q_{0}\right)$
becomes unstable. All the localized states are found below the threshold
for which $G<G_{th}$ or equivalently $g<1$. We note that in the
representation given in Eqs.~(\ref{eq:Rosa1},\ref{eq:Rosa2}) in
which the threshold is automatically unity, the only parameters that
appear are $\left(g,q,\alpha,\beta,s\right)$ as the cavity losses
$\kappa$ have been factored out. Note however that Eqs.~(\ref{eq:VTJ1}-\ref{eq:VTJ3})
are obtained in the limit of small gain and losses. A too strong departure
from the good cavity limit would necessitate larger gain which would induce additional
nonlinearities. Here, the losses could not be factored out anymore.

Interestingly, the equation (\ref{eq:Rosa1}) governing the dynamics
of the transverse profile is a so-called Rosanov equation \cite{RK-OS-88,VFK-JOB-99}
that is known in the context of static transverse auto-solitons in
bistable interferometer. In these works one assumes a monomode continuous
wave (CW) emission along the longitudinal propagation direction which
allows, via the adiabatic elimination of the material variables, to
find an effective equation for the transverse profile. The nonlinear
function $h\left(P\right)$ would correspond to a static saturated
nonlinearity, i.e. $h\left(P\right)\rightarrow1/(1+P)$. A similar
result can be obtained setting $\partial_{z}G=\text{\ensuremath{\partial}}_{z}Q=0$
in Eqs.~(\ref{eq:VTJ1}-\ref{eq:VTJ3}). However, the adiabatic approximation
of the gain along the propagation direction would be incorrect for
a semiconductor material and the reaction time of the gain is known
to profoundly affect the stability of spatio-temporal structures \cite{CPM-NJP-06}.

We note that we operate in a parameter regime where the function $f\left(P\right)$
possesses two fixed points. One solution is unstable and corresponds
to a lower intensity temporal LS while the stable fixed point corresponds
to a higher intensity LS. As such, the system is not bistable for
the CW solution, preventing the existence of static transverse auto-solitons.
It is however bistable for the amplitude of the temporal LS whose
spatial profile may, for proper parameters, coalesce into a transverse
soliton. In the following, we will call the homogeneous spatial solution
the temporal LS with an uniform spatial profile.

Spatial LSs can be found within the region of bistability of the homogeneous
solution. As such, studying in which conditions the homogeneous solutions
of Eqs.~(\ref{eq:Rosa1},\ref{eq:Rosa2}) develops a hysteresis region
informs on the proper parameters for spatial localization. This analysis
is performed in Section 2 of the Appendix, see in particular 
Fig.~\ref{fig:supersubCW} and Fig.~\ref{fig:scaling_CW}. We
summarize here our main results. The critical value of $q_{c}$ above
which one obtain a sub-critical region as a function of the normalized
gain $g$ reads $q_{c}=1/\left(s_{c}-1\right)$. In presence of a
sub-critical region, the folding point, i.e. the minimal value for
the gain $g_{m}$ for which one can obtain a non-zero solution (see
for instance the lower red circle in Fig.~\ref{fig:LS1d_vs_CW}a))
can be approximated by
\begin{eqnarray}
g_{m} & = & -\frac{W_{-1}\left(-e^{-1-\frac{q}{s}}\right)}{1+q}\label{eq:gmLambert}
\end{eqnarray}
with the Lambert-W function $W_{n}$. The extent of the sub-critical
region in which one may expect spatial LSs is then $g\in\left[g_{m,}1\right]$.
The asymptotics in the limit of large saturation and large absorption
simply read
\begin{equation}
\lim_{s\rightarrow\infty}g_{m}=\frac{1}{1+q}\quad,\quad\lim_{q\rightarrow\infty}g_{m}=\frac{1}{s}.\label{eq:scaling_CW-1}
\end{equation}

Our results indicate that large modulations of the absorber $q$ and
large saturation parameters $s$ of course favor the breadth of a
sub-critical region, in agreement with intuition. However, less intuitive
is that a saturation effect exists and marginal increases of the bistability
domain are found for $s>20$ and $q>2$, see  Fig.~\ref{fig:LS1d_vs_CW} in Section 2
of the appendix for more details on the evolution of the folding point.

\section{Results}

The single LS solutions of Eqs.~(\ref{eq:Rosa1},\,\ref{eq:Rosa2})
can be found in the form 
\begin{equation}
A(u,v,t)=a(u,v)\,e^{-i\omega t}\,,\label{eq:LSSepAnsatz}
\end{equation}
where $a(u,v)$ is a complex amplitude with the localized field intensity
$P=|a|^{2}$ and $\omega$ that represents the carrier frequency of
the solution. Substituting Eq.~\ref{eq:LSSepAnsatz} into Eqs.~(\ref{eq:Rosa1},\,\ref{eq:Rosa2})
we are left searching for unknowns $a$ and $\omega$ of the following
equation 
\begin{equation}
\hspace{-0.75cm}(d+i)\left(\partial_{u}^{2}+\partial_{v}^{2}\right)a+i\omega a+f\left(\left|a\right|^{2}\right)a=0.\label{eq:RosaOmA}
\end{equation}
To directly track the LS solutions of Eq.~(\ref{eq:RosaOmA}) in
parameter space, we make use of pde2path~\cite{dohnal2014pde2path,pde2path},
a numerical pseudo-arc-length bifurcation and continuation package
for systems of elliptic partial differential equations. Details regarding
the numerical implementation of the problem can be found in Section 3 of the appendix.

\paragraph*{Evolution of the folding point of the LS solution }

\begin{figure}[ht!]
\begin{tabular}{ll}
(a)  & (b)\tabularnewline
\includegraphics[width=0.45\columnwidth]{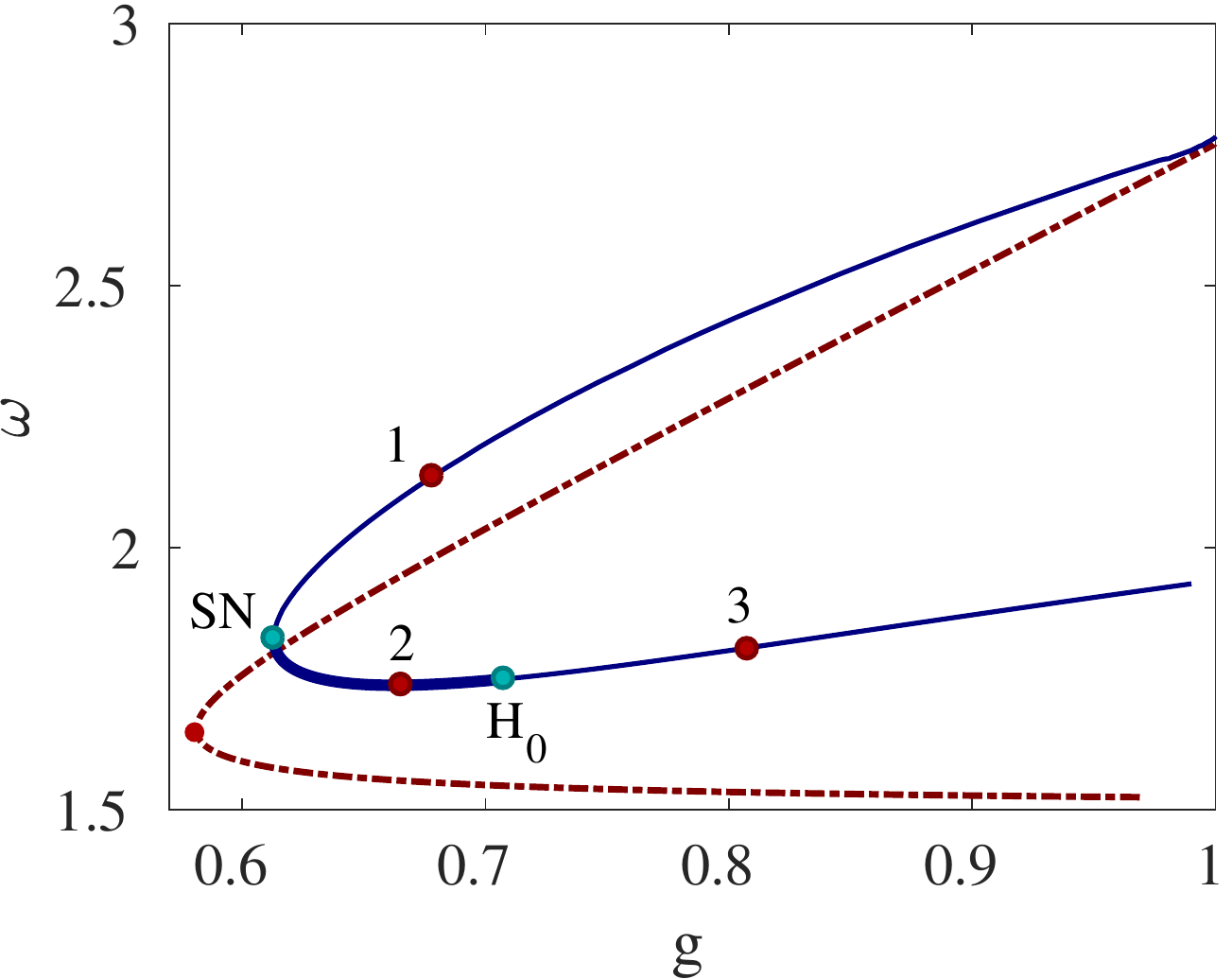}  & \includegraphics[width=0.45\columnwidth]{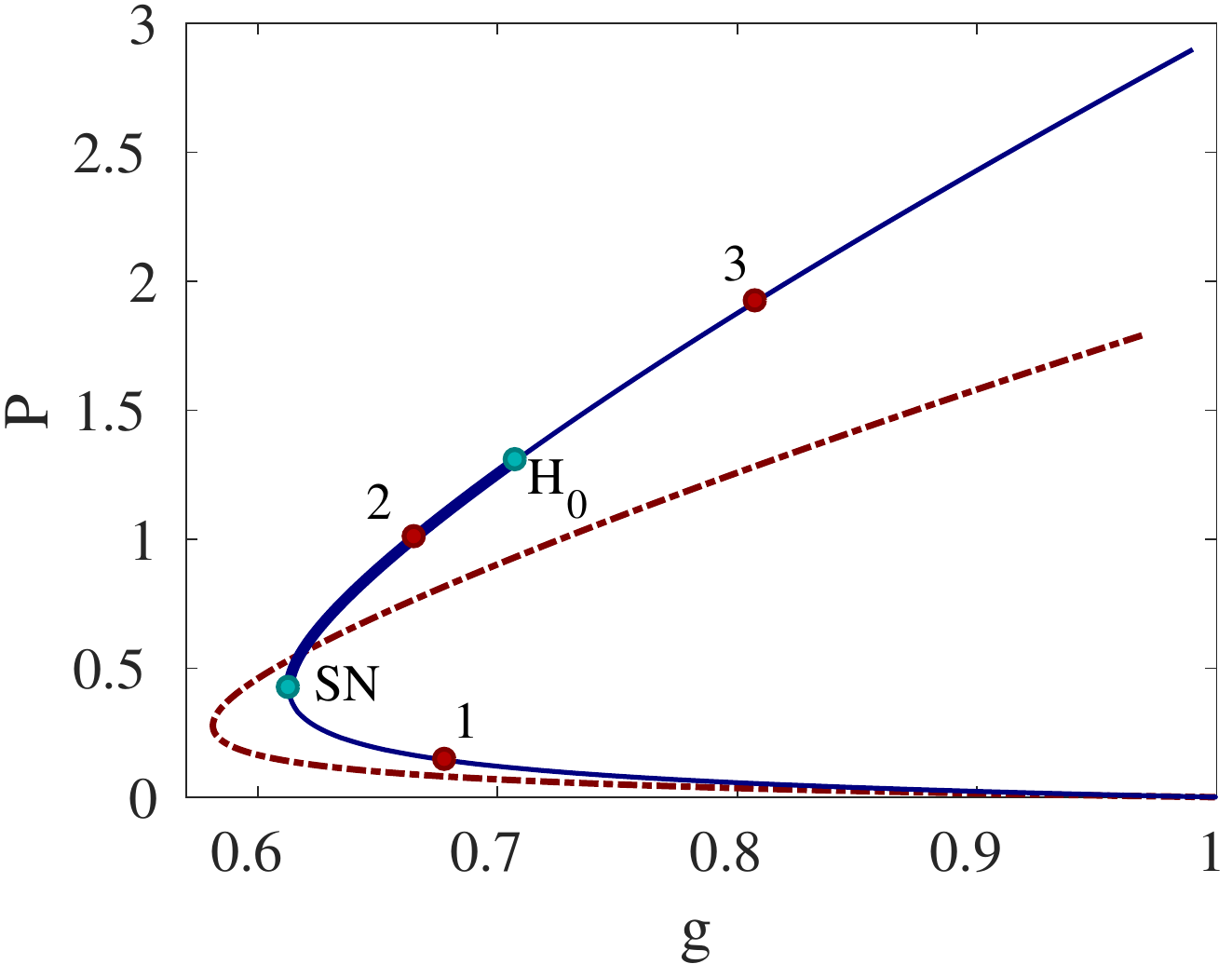}\tabularnewline
(c)  & (d)\tabularnewline
\includegraphics[width=0.45\columnwidth]{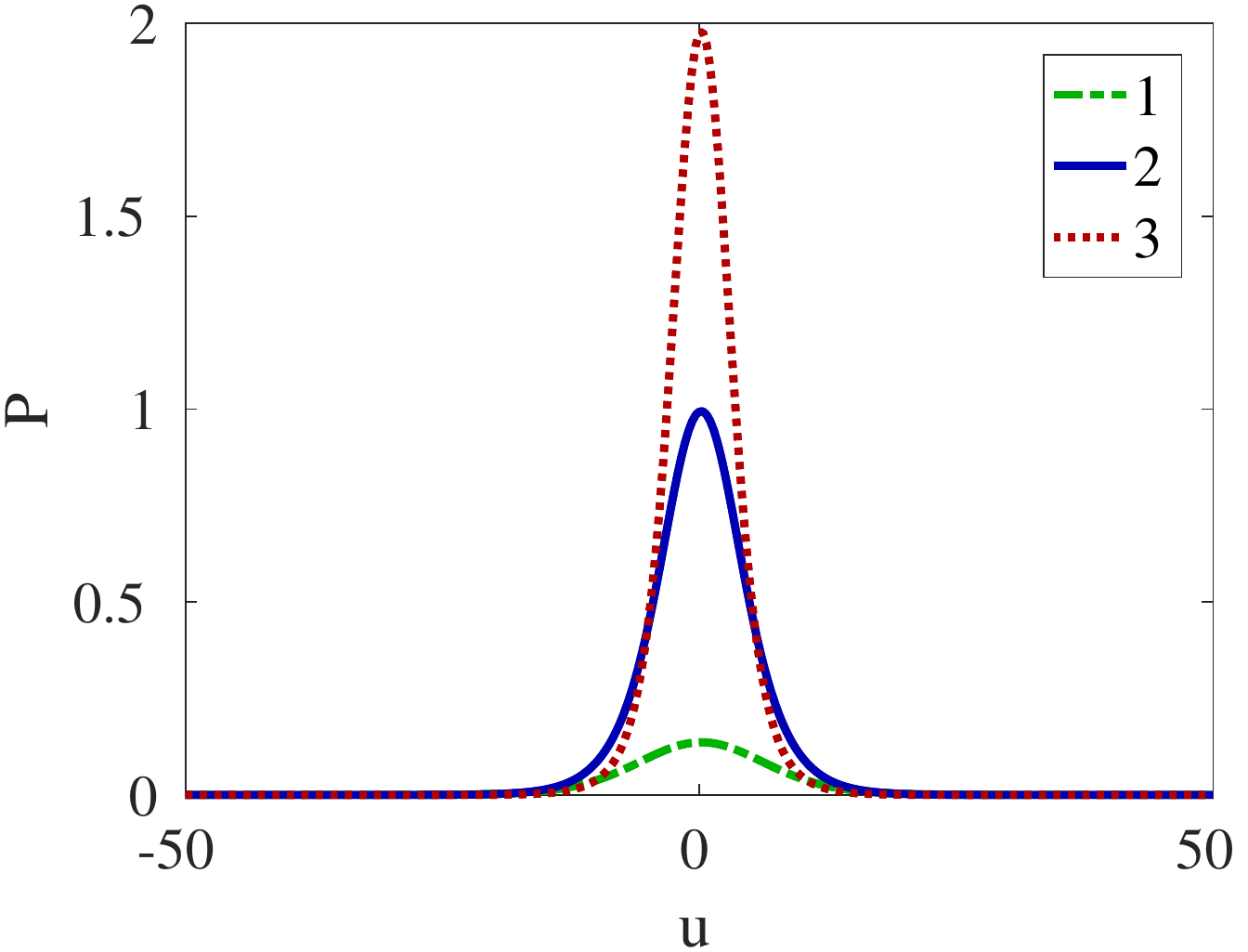}  & \includegraphics[width=0.45\columnwidth]{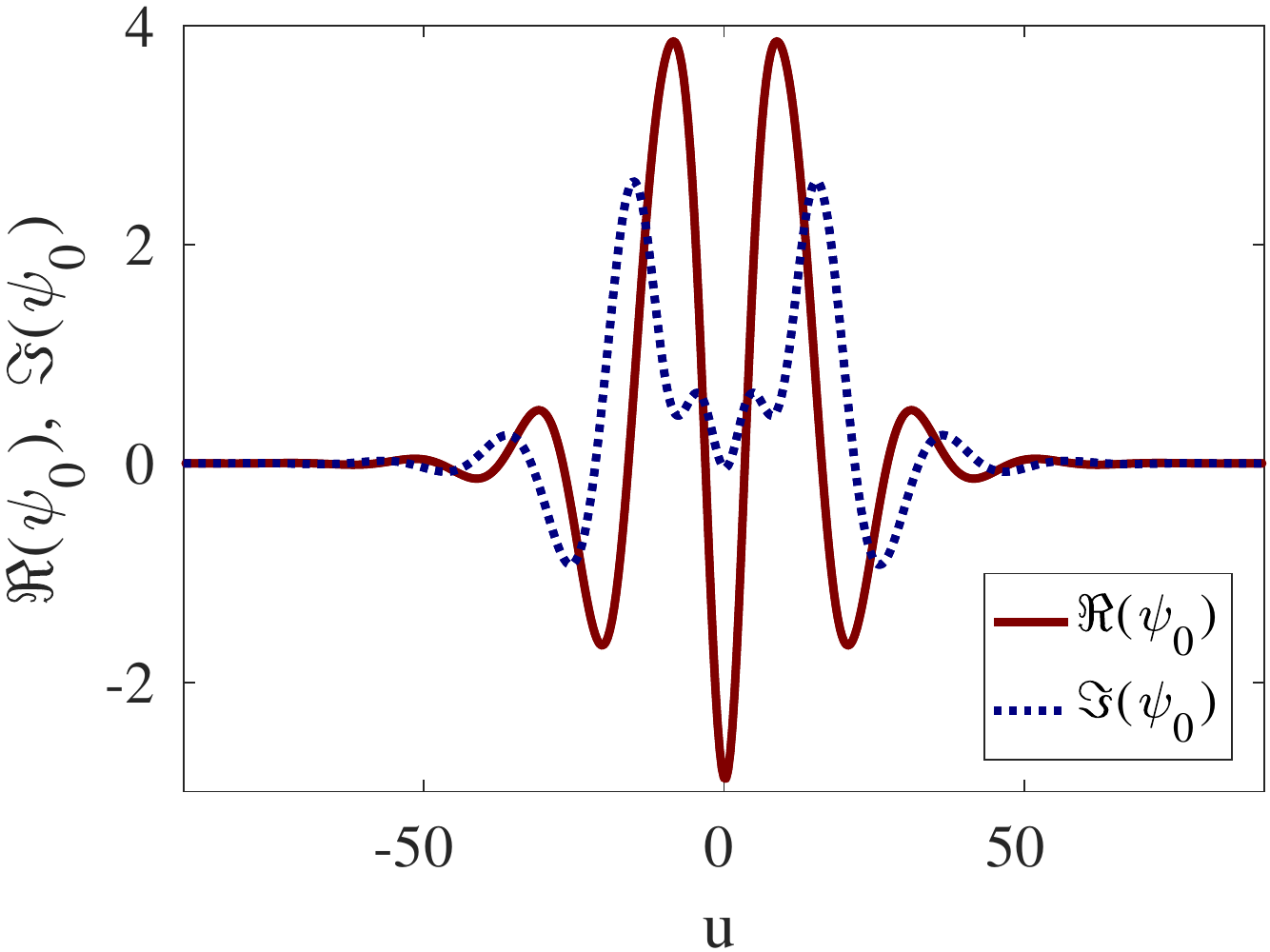} \tabularnewline
\end{tabular}\caption{(Color online) Comparison between the behavior of the branch of the
single LS solution (solid blue line) with that of the homogeneous
solution (red dash-dotted line) calculated for $\alpha=1.5$, $\beta=0.5$.
The evolution of (a) the spectral parameter $\omega$ and (b) the
(peak) intensity $P$ as a function of the normalized gain $g$ is
presented. Three exemplary stationary LS profiles existing for different
values of $g$ are depicted in (c). The LS is stable between the saddle-node
bifurcation point SN and the Andronov-Hopf bifurcation point $\mathrm{H_{0}}$
(thick blue line). We show the real (solid red line) and imaginary
(dashed blue line) parts of the corresponding critical eigenfunction
$\psi_{0}$ in (d). Other parameters are: $q=1.27$, $s=30$, $d=0.01$.}
\label{fig:LS1d_vs_CW} 
\end{figure}

As mentioned, the LS branch also possesses a folding point, see for
instance the blue circle denoted SN in Fig.~\ref{fig:LS1d_vs_CW}
which represents the minimal value of the gain at which localized
states can be obtained. The analysis of the primary folding point
of the LS branch $g_{SN}$ as a function of the normalized absorption
$q$ and the saturation parameter $s$ is detailed in Section 3
of the appendix. It is found that $g_{SN}$ follows closely the evolution
of the folding point of the uniform solution $g_{m}$ with $q$ and
$s$, compare Fig.~\ref{fig:scaling_CW} and Fig.~\ref{fig:folding_p2p} 
in Sections 2 and 3 of the appendix, respectively. 
That is, our predictions for the evolution of the
folding point of the homogeneous solution hold also for the LS branch
and our approximate analytical expression can be used as a guideline.
The one-dimensional folding point is always shifted a few percent
toward higher current values as compared to the uniform case, see
Fig.~\ref{fig:LS1d_vs_CW}a,b) and compare the red and blue curves. A
similar shift for the two-dimensional case exists with respect to
the one dimensional situation, see Fig.~\ref{fig:folding_p2p} in Section 3 of the appendix.

\paragraph*{Bifurcation analysis in one dimension}
We now turn our attention to the possible mechanisms of instability
occurring for increased values of the gain and we present in Fig.~\ref{fig:LS1d_vs_CW}
the bifurcation diagram of Eq.~\ref{eq:RosaOmA} as a function
of $g$ calculated for $\alpha=1.5$ and $\beta=0.5$. In particular,
Fig.~\ref{fig:LS1d_vs_CW}~(a) represents the spectral parameter
$\omega$ of the homogeneous solution (red dash-dotted line)
and that of the one-dimensional LS (solid blue line) as a function
of the gain $g$, while the power $P$ as a function of $g$ for both
homogeneous (dash-dotted red line) and LS solutions (solid blue line)
is depicted in Fig.~\ref{fig:LS1d_vs_CW}~(b). In the case of the
LS we plot the peak intensity of the solution. We note that the bistability
range of the single LS solution is contained in that of the homogeneous
one. In addition, in Fig.~\ref{fig:LS1d_vs_CW}~(c) we depict three
exemplary stationary LS profiles existing for different values of
$g$ as indicated by enumerated labels in both panels~(a,b). The
power of the LS changes significantly along the branch, leading to
a formation of narrow peaks of high intensity at the upper power branch.

Apart form the overall information regarding the branch morphology,
the linear stability of a particular LS solution along the branch
can be obtained directly during continuation. The stability analysis
of Eqs.~(\ref{eq:Rosa1},\ref{eq:Rosa2}) reveals the existence
of several neutral eigenvalues, corresponding to the translation,
phase and Galilean invariances (cf. Refs.~\cite{VRFK-QE-97,VRFK-QE-98}
for static auto-solitons). The results of the linear stability analysis
are shown in Fig.~\ref{fig:LS1d_vs_CW}~(a), (b). Here, thick (thin)
blue curves correspond to stable (unstable) LS solutions. The LS stability
domain lies between the saddle-node bifurcation point SN and the Andronov-Hopf
(AH) bifurcation point $\mathrm{H_{0}}$. There, the branch of LS
gets destabilized via symmetric oscillations, i.e. a breathing of
the LS profile, see \href{https://www.dropbox.com/s/son8gv0njxdkpl4/Hopf_Rosanov_1D_H0.avi?dl=0}{video1}.
An example of the real (solid red line) and imaginary (dashed blue
line) parts of the critical eigenfunction $\psi_{0}$ associated to
this breathing instability is shown in Fig.~\ref{fig:LS1d_vs_CW}~(d).

We note that it was shown in~\cite{J-PRL-16} that, similar to the
case of static autosolitons~\cite{VRFK-QE-97,VFK-JOB-99}, the branch
of one-dimensional LS forms a spiral in the $(g,\omega)$ plane for
vanishing linewidth enhancement factors of both gain and absorber
sections. We show here that, for more realistic values of $\left(\alpha,\beta\right)$,
one finds a simpler branch structure that possesses a single fold.
Our analysis of the evolution of the spiral structure of the branch
and of the spectral parameter $\omega$ as a function of the gain
$g$ for small values of $\left(\alpha,\beta\right)$ can be found
in Section~3, Fig.~\ref{fig:unfolding_spiral} of the appendix. 
We also note that for these more realistic values
of $\left(\alpha,\beta\right)$ the breadth of the stable region is
more extended $\Delta g=g_{H_{0}}-g_{SN}\sim0.1$ while $\Delta g\sim0.03$
in~\cite{J-PRL-16}. This results indicates that the range of stable
LB existence can be widely improved by a proper design of the experimental
devices.

\paragraph*{Bifurcation analysis in two dimensions}

\begin{figure}[ht!]
\begin{tabular}{ll}
(a)  & (b)\tabularnewline
\includegraphics[width=0.45\columnwidth]{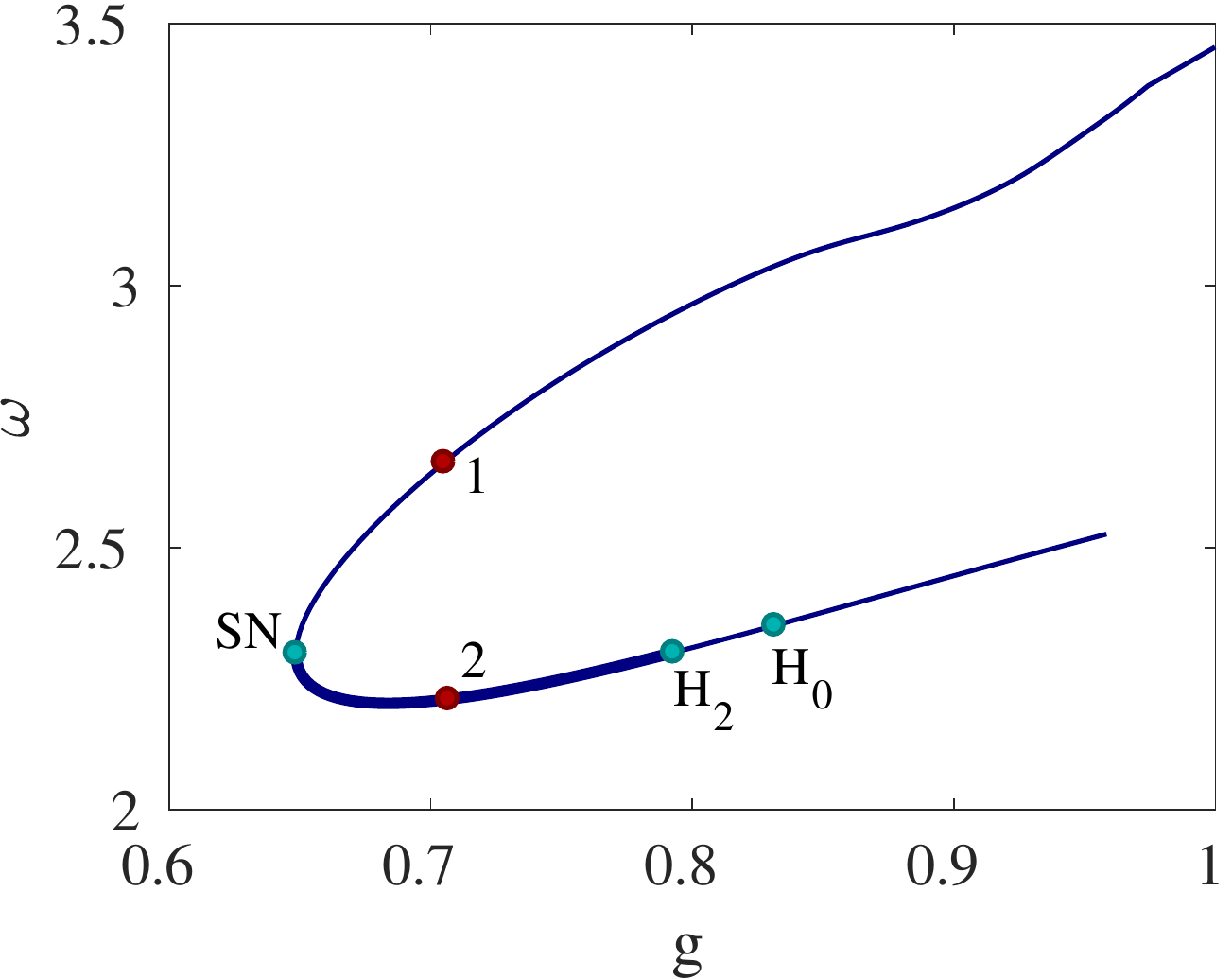}  & \includegraphics[width=0.48\columnwidth]{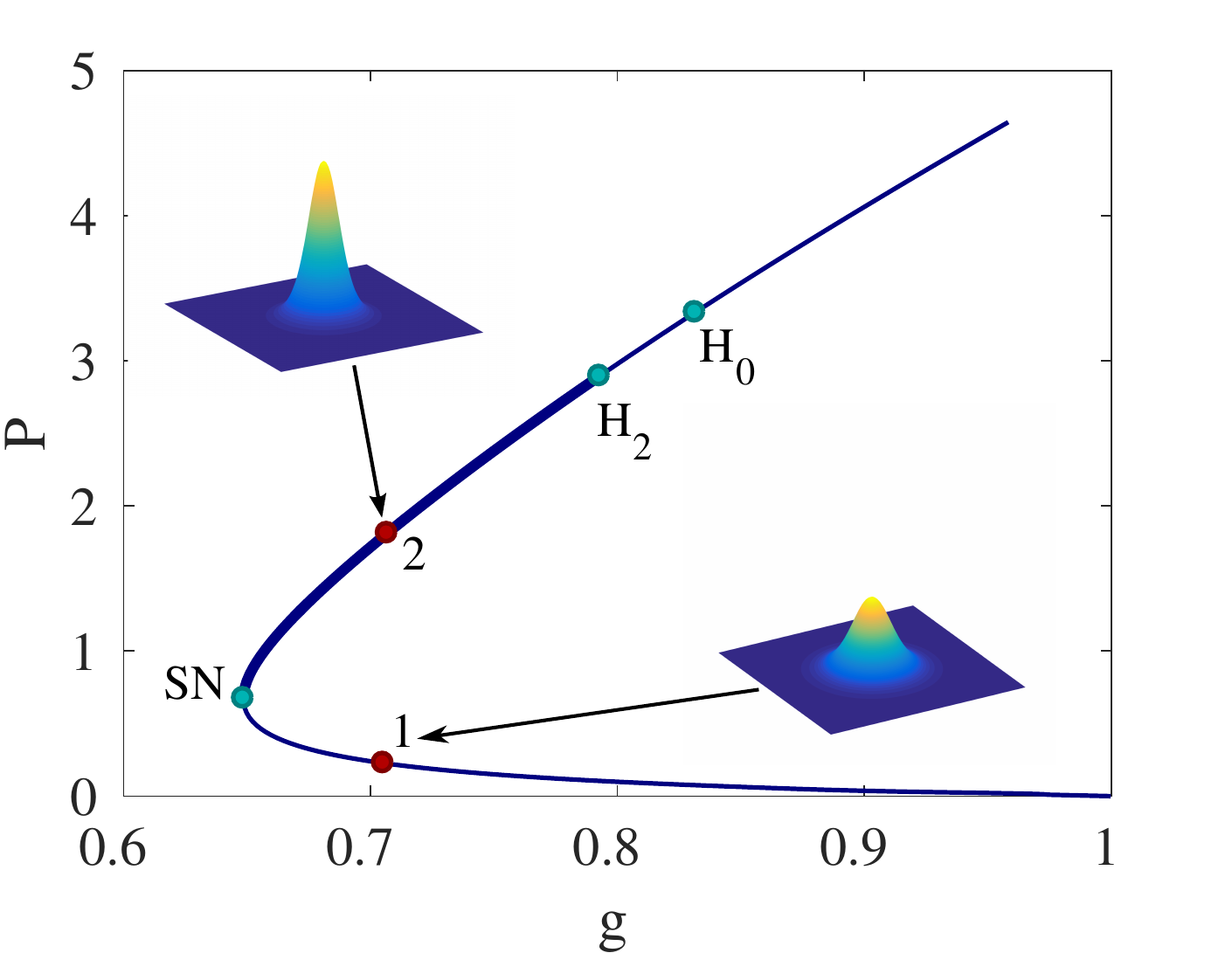}\tabularnewline
(c)  & (d)\tabularnewline
\includegraphics[width=0.45\columnwidth]{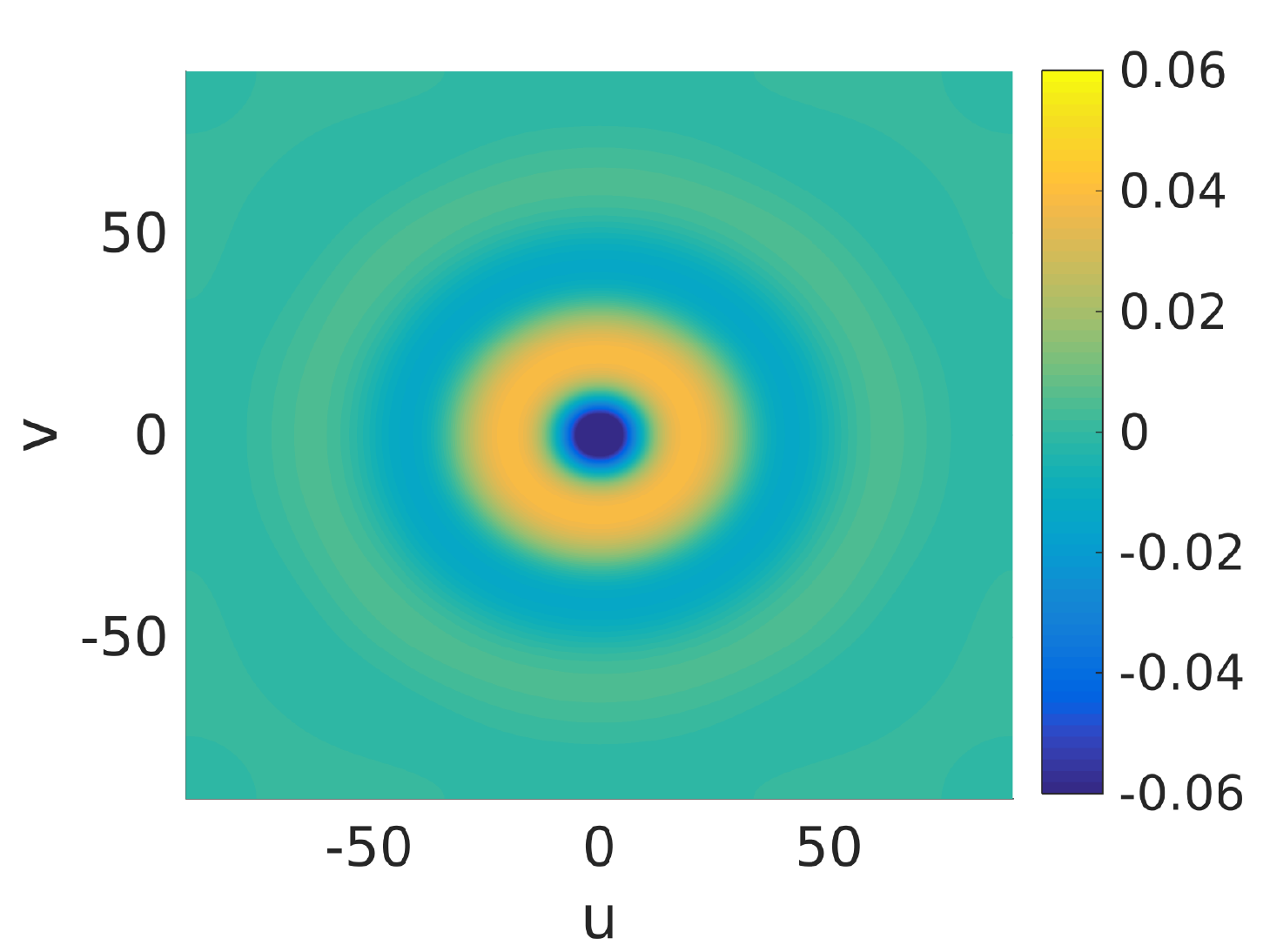}  & \includegraphics[width=0.45\columnwidth]{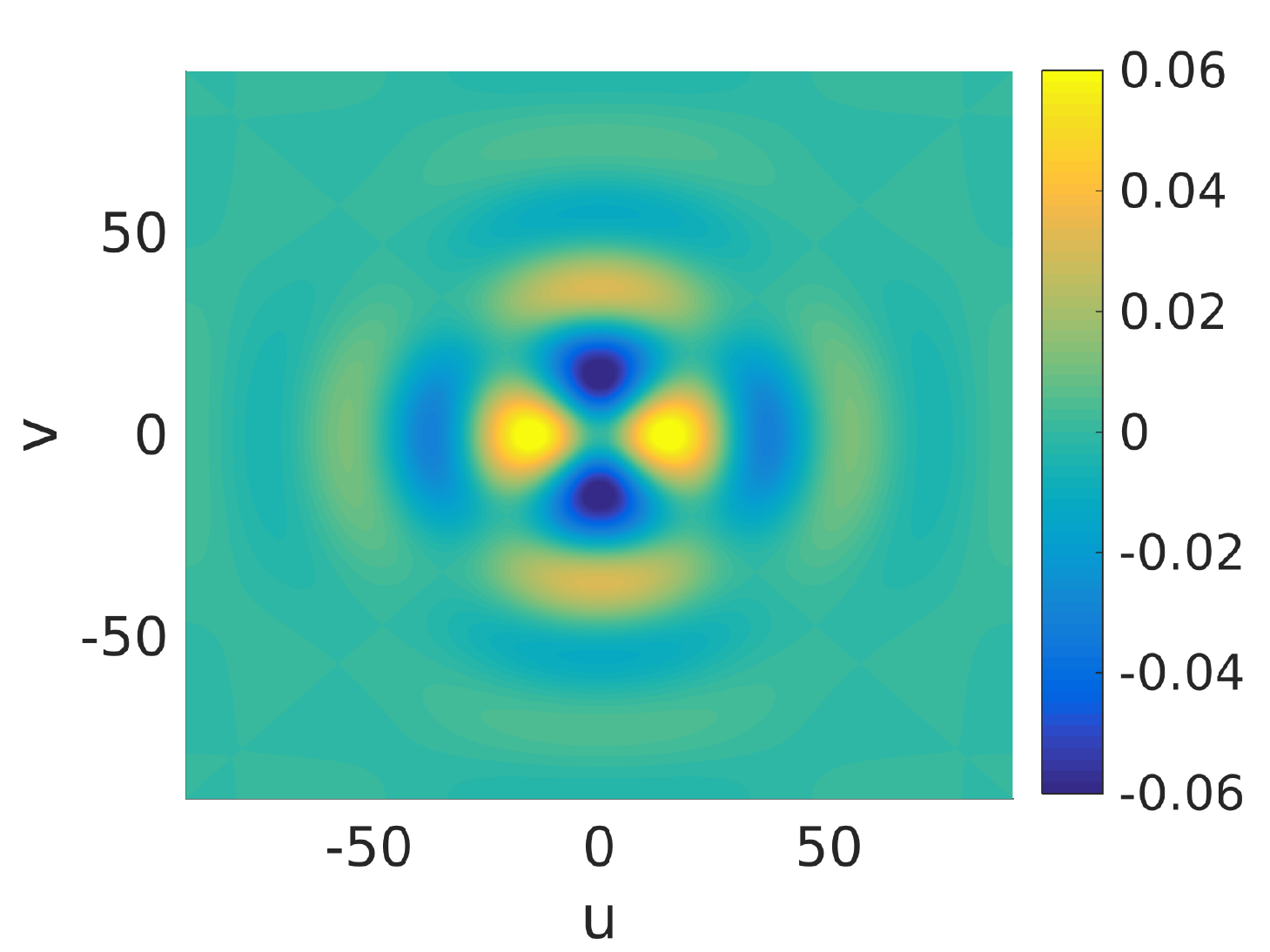} \tabularnewline
\end{tabular}\caption{(Color online) Bifurcation diagram for a two-dimensional LS obtained
for $\alpha=1.7$, $\beta=0.5$ as a function of the gain $g$, showing
the evolution of (a) the spectral parameter $\omega$ while the peak
intensity $P$ is depicted in (b). The LS is stable between the saddle-node
bifurcation point SN and the AH bifurcation point $\mathrm{H_{2}}$
(thick blue line). The secondary AH bifurcation point is indicated
as $\mathrm{H_{0}}$. Two insets in Fig.~\ref{fig:branch_2d}~(b)
show intensity profiles, corresponding to an unstable (label 1) and
a stable (label 2) LS solution obtained at the same gain value $g=0.705$.
We show the real parts of both $n=0$ and $n=2$ critical modes calculated
at $g=0.68$ in (c), (d), respectively. Other parameters are the same
as in Fig.~ \ref{fig:LS1d_vs_CW}.}
\label{fig:branch_2d} 
\end{figure}

We performed a similar analysis in two transverse dimensions obtained
for fixed values of $\alpha=1.7$ and $\beta=0.5$. Our results are
presented in Fig.~\ref{fig:branch_2d}, where the dependence of evolution
of the spectral parameter $\omega$ (panel (a)) and the peak intensity
$P$ (panel (b)) on the gain $g$ is shown. Note that the overall
morphology of the two-dimensional LS branch resembles the one-dimensional
behavior (cf. Fig.~\ref{fig:LS1d_vs_CW}). However, it turns out
that a richer dynamics occurs and that additional modes of instability
are found in two transverse dimensions. Besides a symmetrical radiation
mode, see Fig.~\ref{fig:branch_2d}~(c) leading to an Andronov-Hopf
bifurcation, similar to the one dimensional case and also noted H$_{0}$,
compression-elongation oscillations in the two orthogonal directions
are also possible, see Fig.~\ref{fig:branch_2d}~(d), which correspond
to a bifurcation point noted H$_{2}$. By defining the polar angle
$\phi$ in the transverse plane, we can summarize the situation by
saying that the point spectrum of the two-dimensional LS contains
modes $\propto e^{in\phi}$ with $n=0,\,\pm2$. The mode with $n=0$
results in a symmetrical change of the size of the LS and $n=\pm2$
correspond to a deformation of the LS in two perpendicular directions.
Figure~\ref{fig:branch_2d}~(c), (d) show real parts of both $n=0$
(panel (c)) and $n=2$ (panel (d)) critical modes.

Our linear stability analysis indicates that similar to the one-dimensional
case, the LS solution appears in a saddle-node bifurcation (SN) at
low gain while at high current the dominant mechanism of instability
consists in an AH bifurcation at the point $\mathrm{H_{2}}$, where
the corresponding $n=\pm2$ modes become unstable, see \href{https://www.dropbox.com/s/ifljw4oupxakdgb/Hopf_Rosanov_2D_H2.avi?dl=0}{video2}.
That is, thick (thin) blue curve in Fig.~\ref{fig:branch_2d}~(a),
(b) corresponds to stable (unstable) LS. In addition, two insets in
Fig.~\ref{fig:branch_2d}~(b) show intensity distributions, corresponding
to an unstable (label 1) and a stable (label 2) part of the LS branch
obtained at the same gain value $g=0.705$. Finally, the AH bifurcation
point $\mathrm{H_{0}}$ indicates a secondary instability of the LS
w.~r.~t. modes with $n=0$, see \href{https://www.dropbox.com/s/ttdoexsh08jbxun/Hopf_Rosanov_2D_H0.avi?dl=0}{video3}.
Here, additional symmetrical oscillations of the LS shape are expected.
Notice that the order of both AH bifurcations strongly depends on
the linewidth enhancement factors of both gain and absorber sections. 

\paragraph{Range of existence of the single LS.}

In \cite{J-PRL-16}, the linewidth enhancement factors of both the
gain and absorber sections were set $\alpha=\beta=0$ as a demonstration
that the carrier induced self-focusing effects played no role in the
LB formation and that the confinement mechanism was different than
the one found in conservative systems as, e.g., in \cite{S-OL-90}.
In this section, we study the influence of the $\alpha$ and $\beta$
factors that are set to more typical values and we find that an extended
range of stability exists by mapping the position of the points H$_{0}$
and H$_{2}$ limiting the existence of stable LBs at high current.
Adding to these results the evolution of the folding point SN, allows
us to disclose the range of existence of stable LSs in one and two
transverse spatial dimensions.

\begin{figure}[ht!]
\begin{tabular}{ll}
(a)  & (b)\tabularnewline
\includegraphics[width=0.45\columnwidth]{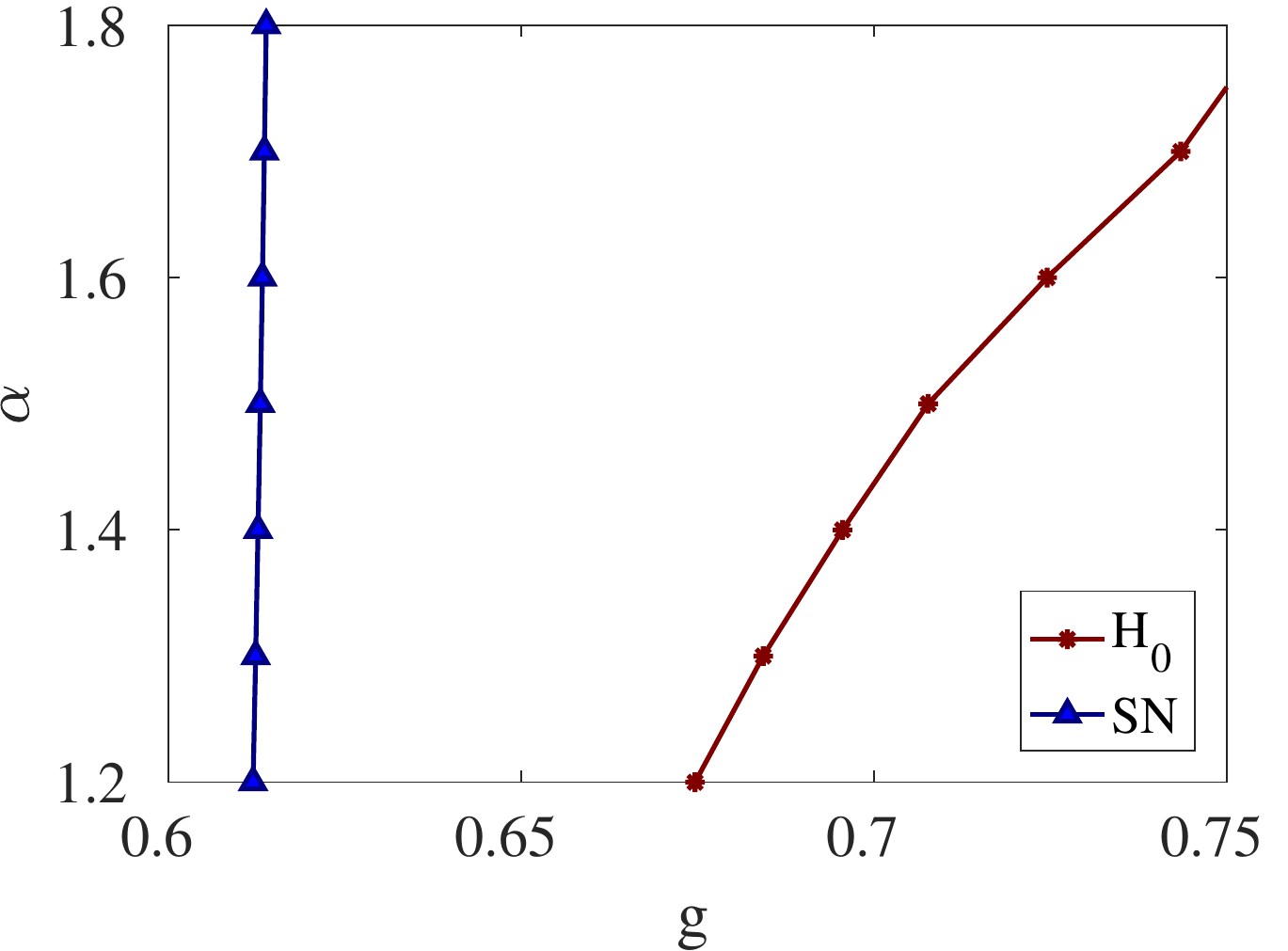}  & \includegraphics[width=0.45\columnwidth]{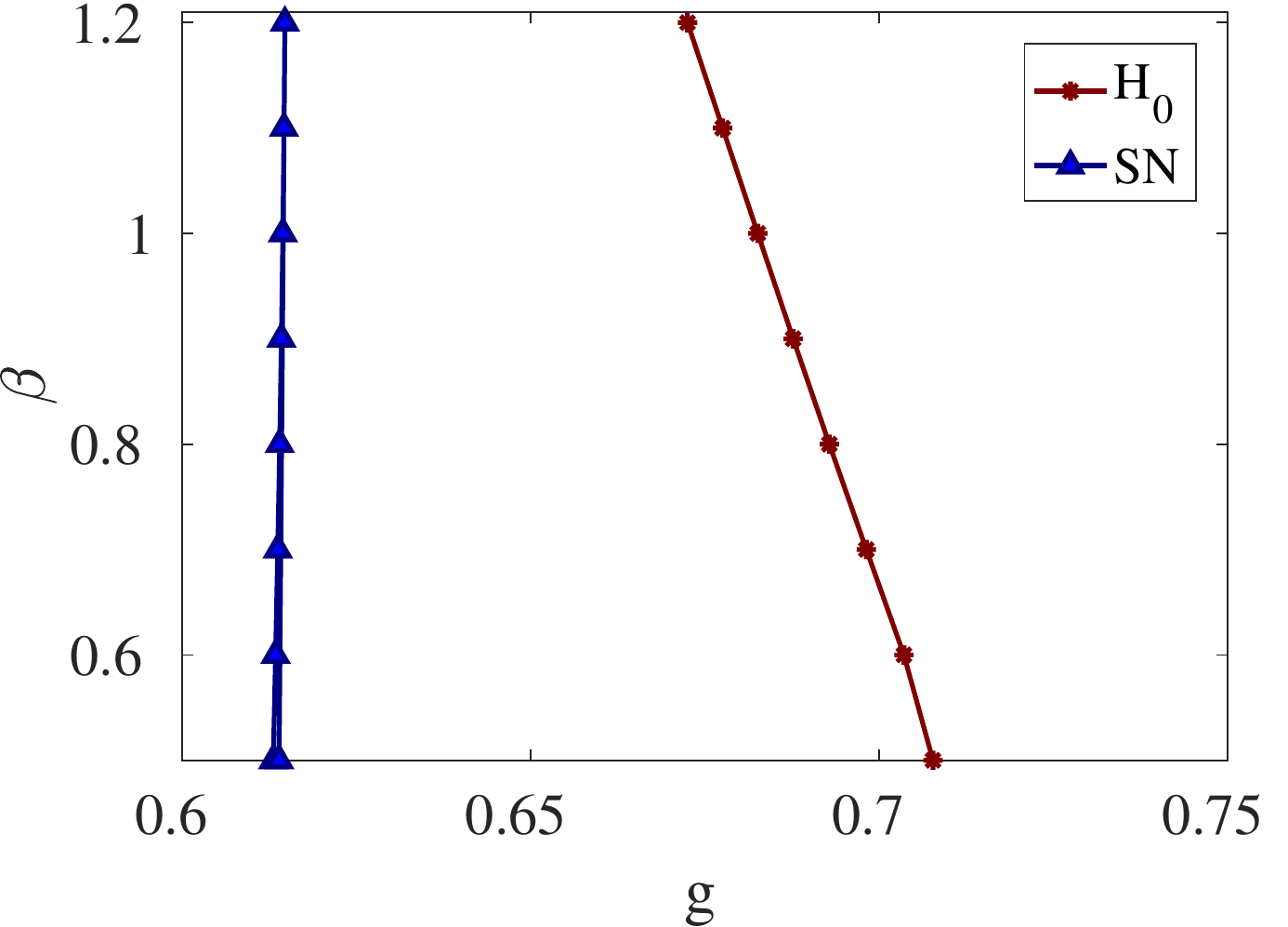}\tabularnewline
(c)  & (d)\tabularnewline
\includegraphics[width=0.45\columnwidth]{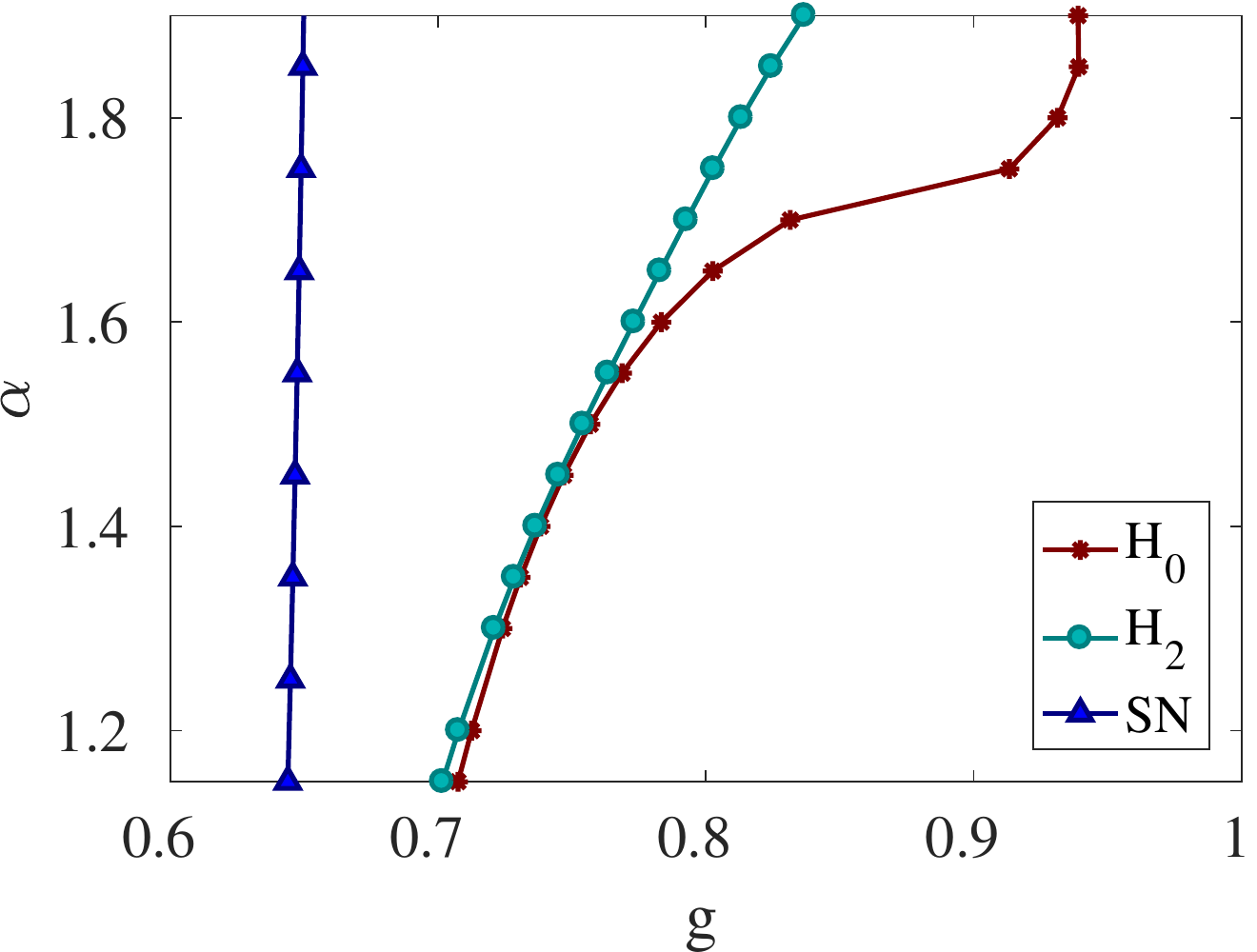}  & \includegraphics[width=0.45\columnwidth]{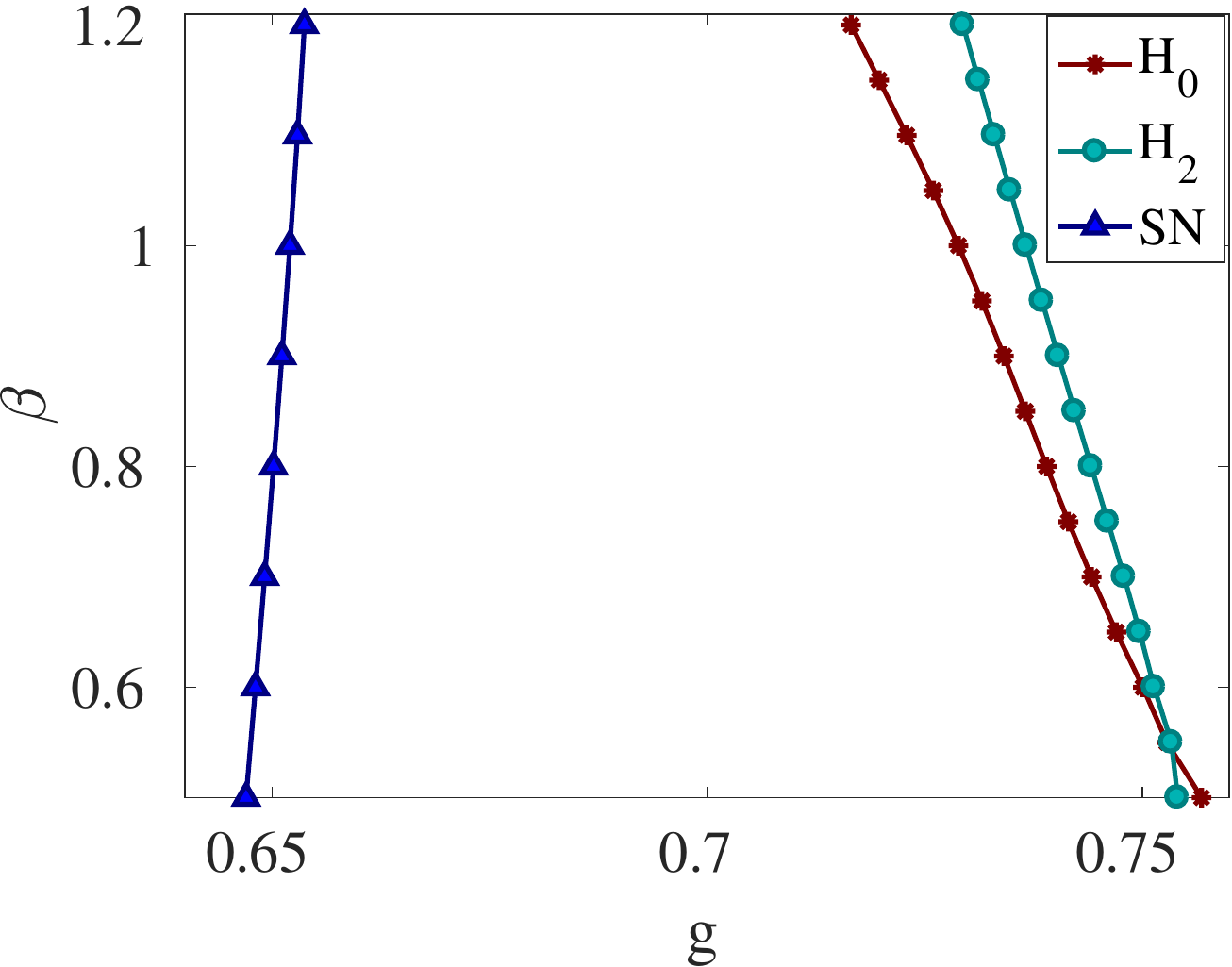}\tabularnewline
\end{tabular}\caption{(Color online) (a), (b). One-dimensional stability diagrams showing
the threshold for the fold (blue triangles) and $n=0$ AH bifurcations
(red stars) in terms of the dependence of $\alpha$ on the gain $g$
for (a) $\beta=0.5$ and (b) $\alpha=1.5$. (c),(d). The same stability
diagram in two dimensions. Cyan circles depict the Hopf line corresponding
to the $n=2$ instability. The range of stability grows with $\alpha$
and shrinks with $\beta$.}
\label{fig:bif_diag_2d} 
\end{figure}

As we mentioned above, in the one-dimensional case, at high current
the dominant mechanism of instability consists in an Andronov-Hopf
bifurcation H$_{n}$ with $n=0$, whereas in the two-dimensional case
we identified several modes of destabilization, where the amplitude
of the LS oscillates uniformly in space ($n=0$) and another where
the breadth of the LS breathes ($n=2$). In order to study the influence
of the $\alpha$ and $\beta$ factors on the stability range of the
single LS solution we perform a two parameter continuation of the
fold and AH bifurcation points as a function of the gain $g$ and
the linewidth enhancement factors of the gain and the absorber sections.
Our results are depicted in Fig.~\ref{fig:bif_diag_2d}.

Figure~\ref{fig:bif_diag_2d} panels (a), (c) represent the stability
diagrams in the $(g,\alpha)$ plane for the fixed value of $\beta=0.5$
for one and two spatial dimensions, respectively. Here, blue triangles
indicate the fold threshold, whereas red stars stand for the boundary
of the AH bifurcation with $n=0$. In addition, in Fig.~\ref{fig:bif_diag_2d}~(c)
cyan circles depict the second AH line $n=2$. Note that depending
on the relation between $\alpha$ and $\beta$ one of these two oscillation
modes can govern the primary instability threshold. Our results reveal
that in both one and two dimensions the range of the stability increases
toward higher $\alpha$ values. Although the fold position keeps almost
constant for increasing $\alpha$, both AH lines move toward higher
currents. However, the width of the stability region strongly depends
on $\beta$ as shown in Fig.~\ref{fig:bif_diag_2d}~(b), (d), where
the stability diagram for increasing $\beta$ and a fixed moderate
value of $\alpha=1.5$ is presented for both one- and two-dimensional
continuations. Here, in both cases the range of stability decreases
for growing $\beta$.

\begin{figure}[ht!]
\begin{tabular}{lll}
(a) & (b) & (c)\tabularnewline
\includegraphics[width=0.3\columnwidth]{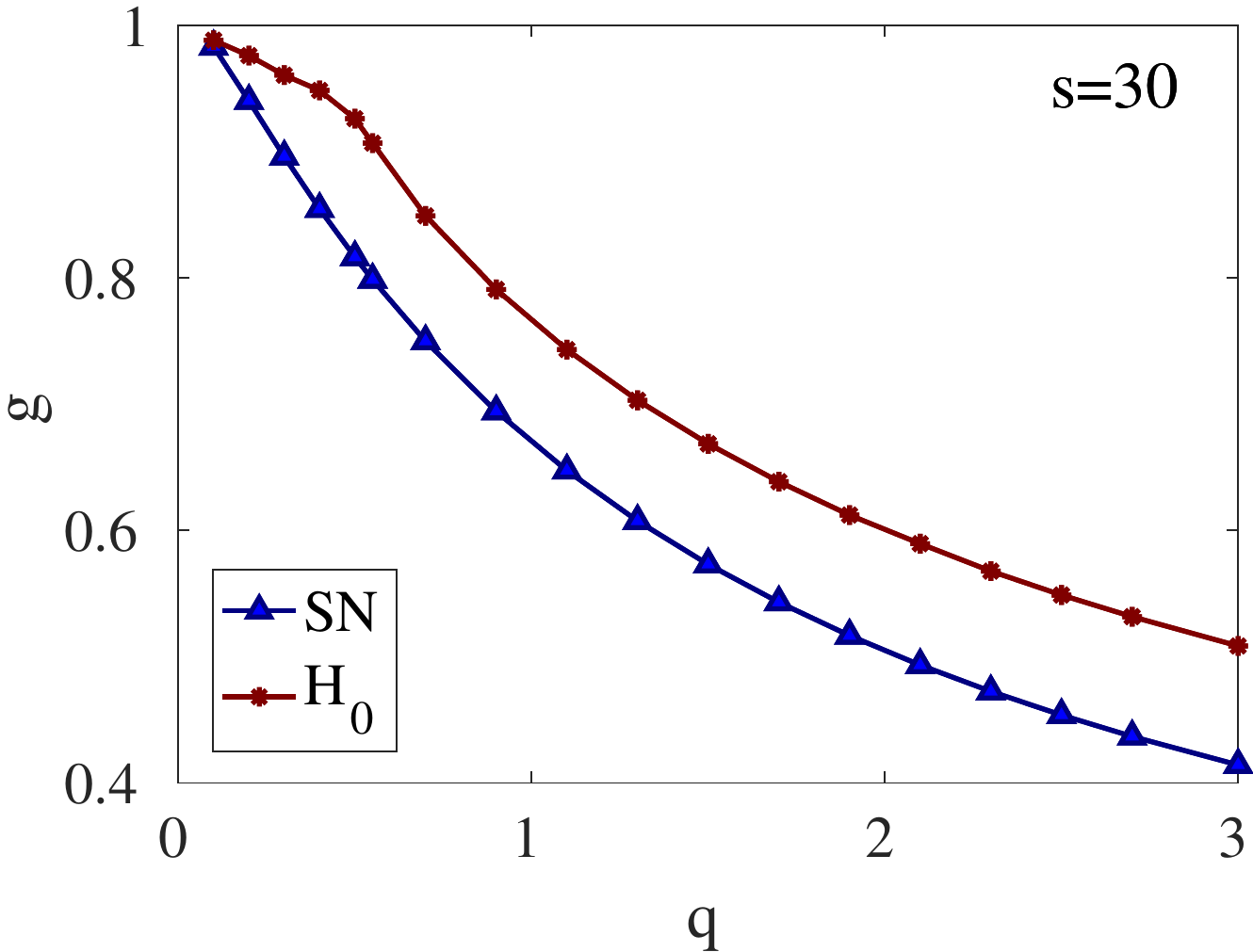} & \includegraphics[width=0.3\columnwidth]{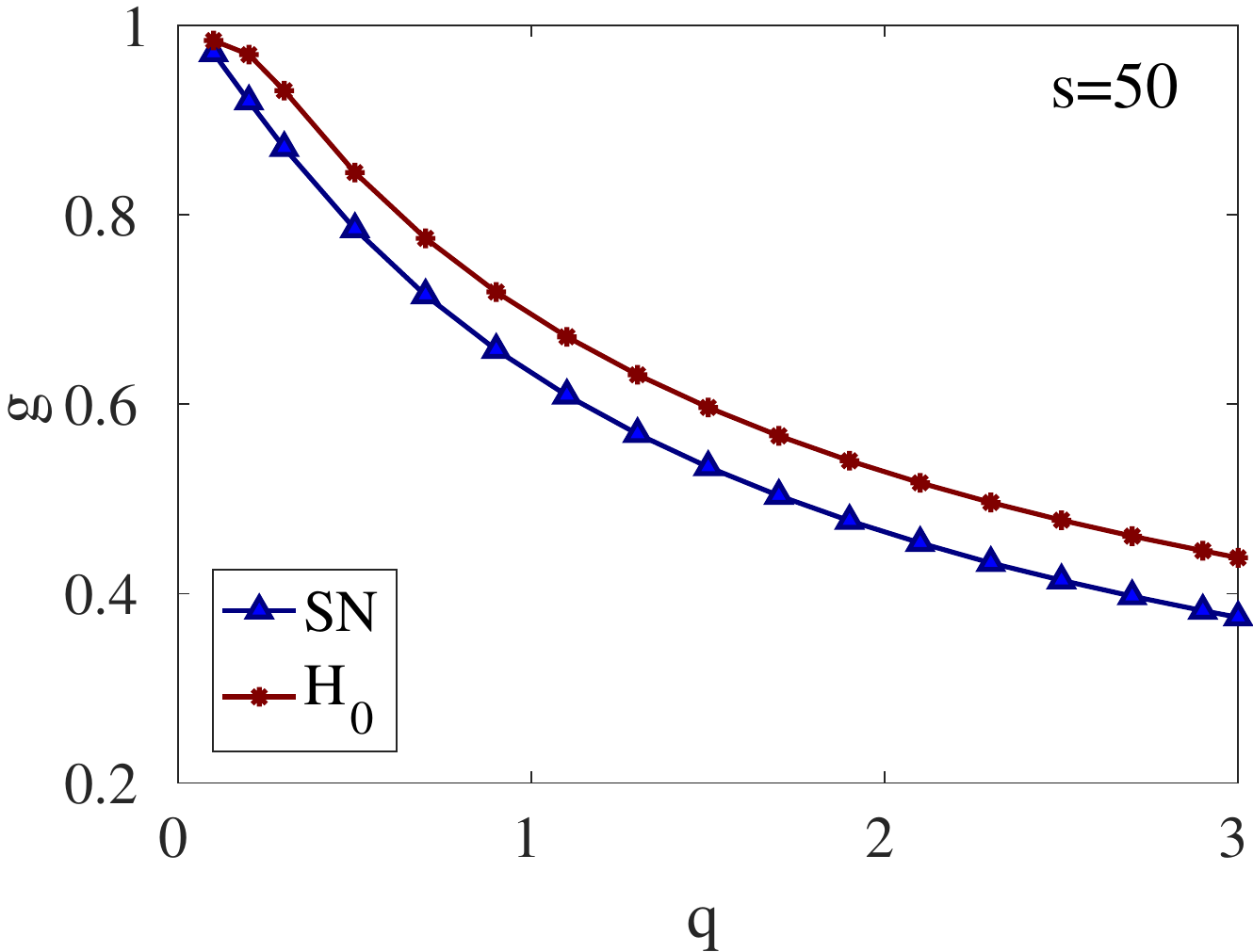} & \includegraphics[width=0.3\columnwidth]{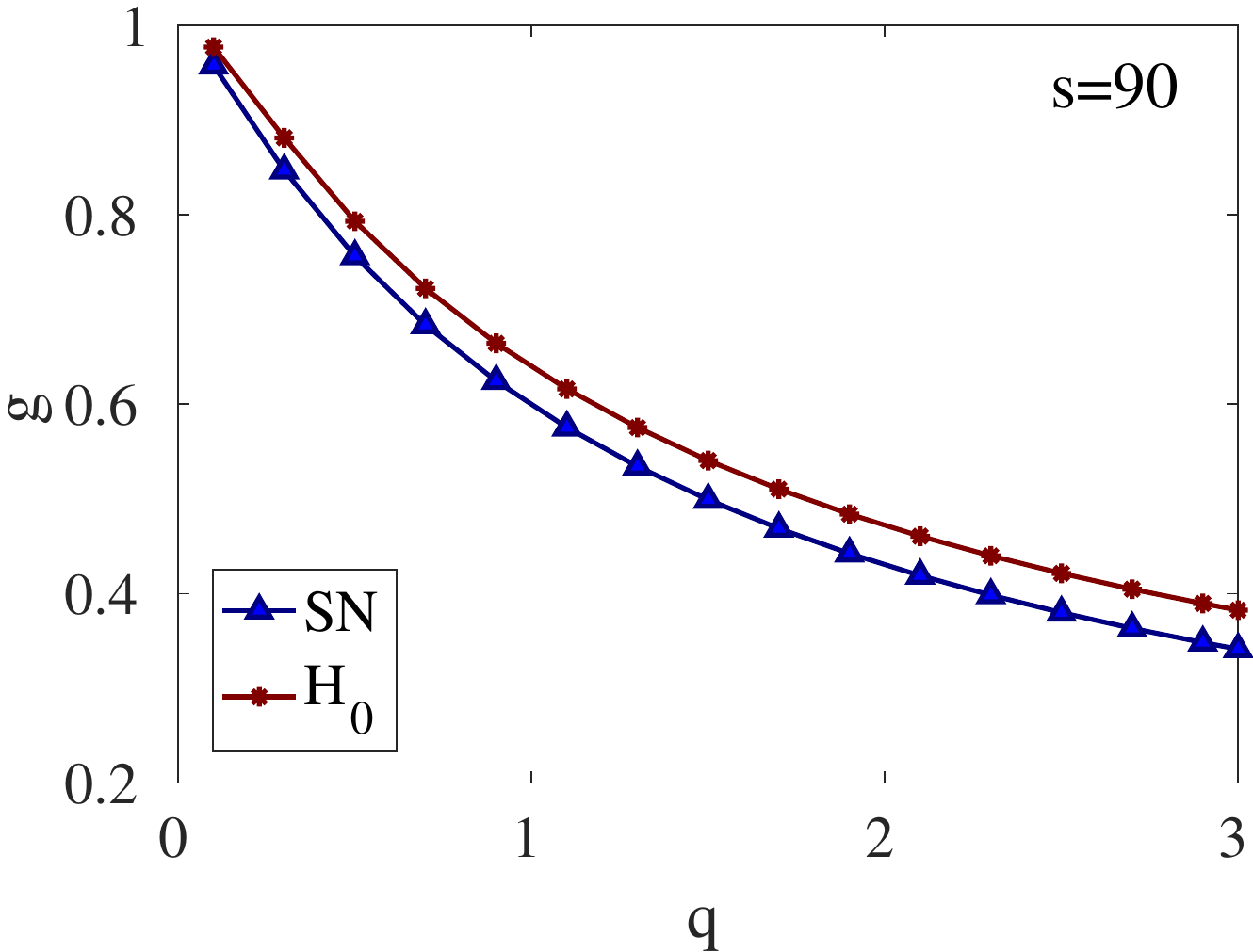}\tabularnewline
\end{tabular}\caption{(Color online). One-dimensional stability diagrams showing the threshold
for the fold (blue triangles) and the $n=0$ AH bifurcations (red
stars) in terms of the dependence of $s$ on the gain $g$ for $\left(\alpha,\beta\right)=\left(1.5,0.5\right)$.The
range of stability shrinks for too larges values of $s$. Similar
trends are found in in two dimensions.}
\label{fig:bif_diag_1d_s} 
\end{figure}

Finally, we performed a similar analysis as a function of the saturation
parameter $s$ as shown in Fig.~\ref{fig:bif_diag_1d_s}. One notices
that for generic values of $\left(\alpha,\beta\right)=\left(1.5,0.5\right)$,
the stability region \emph{decreases} if the saturation parameter
$s$ gets too large, which is a counter-intuitive result.

\paragraph{Numerical Simulations of the Haus model}

By using the guidelines obtained from the analysis of the homogeneous
and the transverses solutions, we now turn our attention to the predictions
obtained by directly integrating Eqs.~(\ref{eq:VTJ1}-\ref{eq:VTJ3}),
the details of the numerical methods are depicted in Section~4
of the appendix. Our results are summarized in Fig.~\ref{fig:bif_diag_Haus_2d_gab}
in which we plot the energy of the LB $\mathcal{E}=\iint|E|^{2}dxdz$
and depict the region of stable existence as a function of the gain
$g$ and the linewidth enhancement factors $\alpha$ and $\beta$.
Nicely, the dominant trends predicted by analyzing the effective equation
of the transverse profile Eq.~(\ref{eq:Rosa1}) are confirmed, a
strong increase of the stable region with $\alpha$, and a moderate
decreases for increasing $\beta$. In \href{https://www.dropbox.com/s/wxslw73ie42x72q/Hopf_Haus_2D_H0.avi?dl=0}{video4},
we depict the instability mechanisms in 2D that corresponds to H$_{0}$,
as predicted, while in 3D, the dominant mechanism is governed by the
H$_{2}$ instability \href{https://www.dropbox.com/s/2884z08e1q34793/Hopf_Haus_3D_H2.avi?dl=0}{video5}.

The transverse and longitudinal Full Width at Half Maximum (FWHM),
the carrier frequency $\omega$ and drifting speed $\upsilon$ of
the LBs are presented in Section~4 of the appendix.

\begin{figure}[ht!]
\includegraphics[width=1\columnwidth]{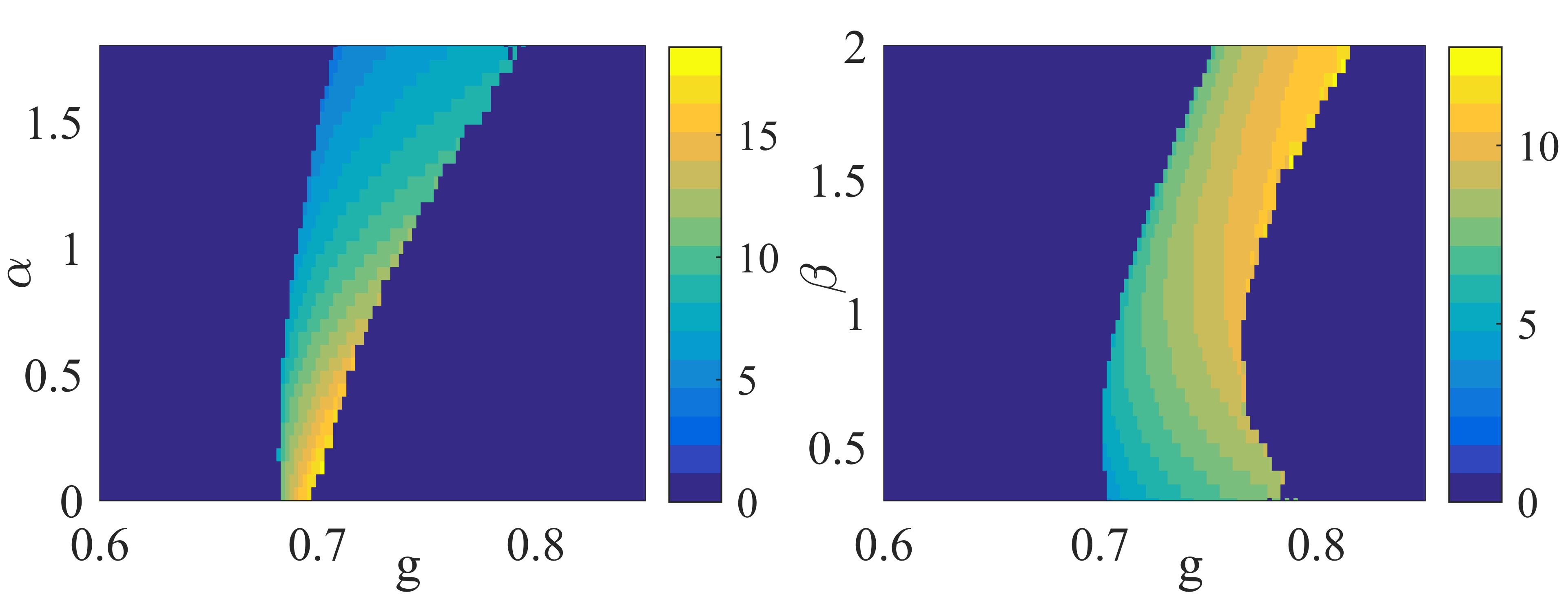}\caption{(Color online). Two-dimensional bifurcation diagram showing the region
of stable existence of the LBs by integrating the two-dimensional
Haus Equation in $\left(x,z\right)$ as a function of the gain $g$
and the parameters $\alpha$ (a) and $\beta$ (b). The color code
represents the energy of the LB $\mathcal{E}$. The range of stability
increases with $\alpha$ and reduces for small values of $\beta$
in agreement with Fig.~\ref{fig:bif_diag_2d}. Similar trends were
found in three dimensions. Parameters are $\left(\gamma,\kappa,\Gamma,Q_{0},s,d\right)=\left(40,0.8,0.04,0.3,30,10^{-2}\right)$
and $\beta=0.5$ (a), $\alpha=1.5$ (b).}
\label{fig:bif_diag_Haus_2d_gab} 
\end{figure}

\begin{figure}[ht!]
\includegraphics[bb=0bp 0bp 6300bp 1580bp,clip,width=1\columnwidth]{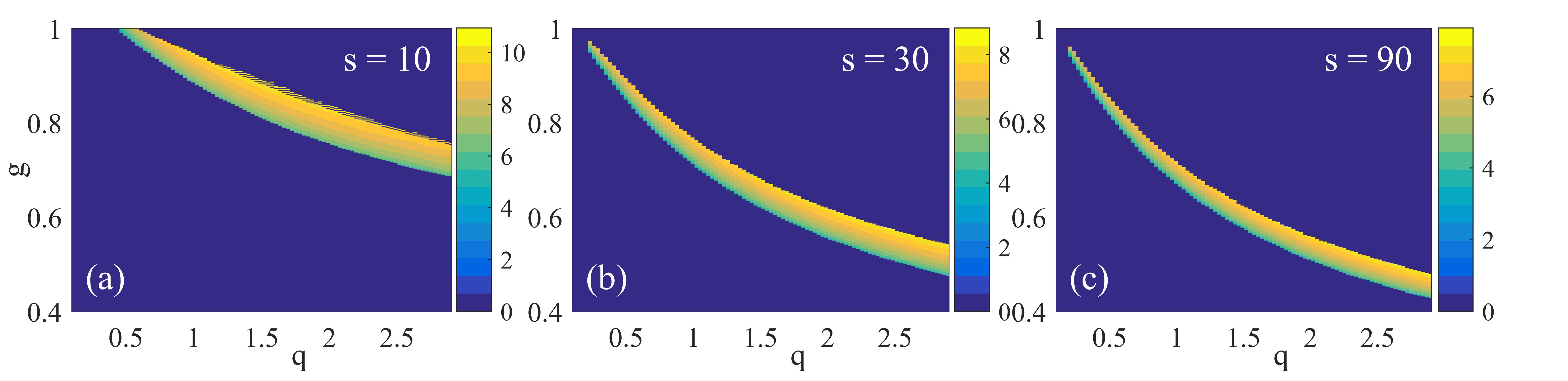}\caption{(Color online). Two-dimensional bifurcation diagram showing the region
of stable existence of the LBs by integrating the two dimensional
Haus Equation as a function of the gain $g$ and the modulation of
the absorption $q$. The color code represents the energy of the LB
$\mathcal{E}$ for increasing values of the saturation parameter $s$.
We find that the range of stability decreases with increasing values
of $s$. Similar trends were found in three dimensions. Parameters
are $\left(\gamma,\kappa,\alpha,\beta,\Gamma,d\right)=\left(40,0.8,1.5,0.5,0.04,10^{-2}\right)$.}
\label{fig:bif_diag_Haus_2d_sgq} 
\end{figure}

A similar behavior as that predicted in Fig.~\ref{fig:bif_diag_1d_s}
can be found while performing the integration of the Haus model in
Eqs.~(\ref{eq:VTJ1}-\ref{eq:VTJ3}) as a function of the saturation
parameter $s$. Our results are summarized in Fig.~\ref{fig:bif_diag_Haus_2d_sgq}
where we represented the energy, the other parameters of the solutions
are depicted in Section~4 of the appendix. 
The dominant trends of a decrease of the region of existence with
increased values of $s$ is confirmed.

\section{Conclusion}

In conclusion, we discussed how the dynamics of 3D light bullets can
be successfully approximated in a wide parameter range by a simplified
model governing the dynamics of the transverse profile. The bifurcation
analysis of this effective model allows to obtain guidelines regarding
the existence and the stability of the LBs found in the full problem.
We have found that, as a function of the gain, the stability range
is governed by the evolution of a lower limit point where the LB solution
ceases to exist and an upper one where the system develops a breathing
instability that can result either in a homogeneous oscillation of
the profile or orthogonal compression-elongation oscillations. We
have found that, contrary to intuition, too large saturation parameters
or modulation of the losses are either detrimental or irrelevant to
the range of stability of the LBs. Finally, direct numerical simulations
performed on a HPC cluster of the full system confirmed the predictions
of the simplified model. In our analysis we have found that the mechanism
of instability of the LBs are essentially those of the transverse
profile. Yet, we believe that additional instabilities may take place
for larger values of $\left(\alpha,\beta\right)$ for which the temporal
LS that is performing the backbone of the spatial soliton can become
unstable. Finally, instabilities that do not pertain to either the
spatial or the temporal degree of freedom but to both simultaneously
can not be ruled out and will be the topic of further studies.

\section*{Acknowledgments}
S.G. acknowledges the Universitat de les Illes Baleares for funding
a stay where part of this work was developed. J.J. acknowledges Joan
Arbona for technical support regarding the UIB cluster Foner, the 
project COMBINA (TEC2015-65212-C3-3-P AEI/FEDER UE) and the Ramón y Cajal fellowship. 
J.J. and S.G. thanks Christian Schelte for a careful reading of the manuscript.

\appendix*

\section{}
\subsection{Derivation of the Effective Rosanov Equation} \label{subsec:Rosanov_Deriv}

We assume that the field reads $E\left(r_{\perp},z,\sigma\right)=A\left(r_{\perp},\sigma\right)p\left(z\right)$
with a short normalized temporal pulse $p\left(z\right)$ of length
$\tau_{p}$ and a slowly evolving amplitude $A\left(r_{\text{\ensuremath{\perp}}},\sigma\right)$.
Next, we use the fact that, during the emission of a LB, the stimulated
terms are dominant in Eqs.(\ref{eq:VTJ2}-\ref{eq:VTJ3}), i.e.
$\left|E\right|^{2}\gg1$ such that 
\begin{eqnarray}
-\int_{0}^{\tau_{p}}G\left|E\right|^{2}dz & \simeq & \int_{0}^{\tau_{p}}\partial_{z}G\,dz=G_{f}-G_{i}
\end{eqnarray}
with $G_{i}$ (resp. $G_{f}$) the gain before (resp. after) the pulse
emission, see \cite{N-JQE-74,VT-PRA-05} for more details. By integrating
Eq.~(2) in the regime $\left|E\right|^{2}\gg1$ we find
that $G\left(z\right)=G\left(0\right)\exp\left(-\int_{0}^{z}\left|p\right|^{2}\left|A\right|^{2}dz'\right)$
and $G_{f}=G_{i}\exp\left(-\left|A\right|^{2}\right)$. Considering
Eq.~(3) with similar arguments one finds $Q_{f}=Q_{i}\exp\left(-s\left|A\right|^{2}\right)$.
Multiplying the field equation Eq.~(1) by $\bar{p}$,
integrating over the pulse length, neglecting the contribution $\gamma^{-1}\dot{p}$,
and using the above expression of the stimulated terms, we find that
the equation governing the dynamics of the transverse profile reads
\begin{eqnarray}
\partial_{\sigma}A & = & (d+i)\Delta_{\perp}A+A\,F\left(\left|A\right|^{2}\right).\label{eq:Rosanov}
\end{eqnarray}
The expression of the nonlinear function $F$ is 
\begin{eqnarray}
\negthickspace\negthickspace\negthickspace\negthickspace\negthickspace F\negthickspace & = & \sqrt{\kappa}\left[1+\frac{1-i\alpha}{2}G_{0}h\left(P\right)-\frac{1-i\beta}{2}Q_{0}h\left(sP\right)\right]-1\label{eq:f}
\end{eqnarray}
with $h\left(P\right)=\left(1-e^{-P}\right)/P$. We replaced in Eq.~(12)
the values of the gain and of the absorption at the beginning of the
pulse by their equilibrium values by taking advantage of the long
cavity limit. It allows us to assume that the gain and absorption
looses entirely their memory at the next round-trip. 

The lasing threshold above which the off solution $\left(E,\,G,\,Q\right)=\left(0,\,G_{0},\,Q_{0}\right)$
becomes unstable is 
\begin{eqnarray}
G_{th} & = & \frac{2}{\sqrt{\kappa}}-2+Q_{0}.
\end{eqnarray}
As we operate in the region below threshold in which the temporal
LSs are bistable with the trivial off solution, we define the gain
normalized to threshold and the normalized absorption $q$ as 
\begin{equation}
g=G_{0}/G_{th}\quad,\quad q=Q_{0}/\left(\frac{2}{\sqrt{\kappa}}-2\right).
\end{equation}
Defining the scaled spatial and temporal coordinates as $t=\left(1-\sqrt{\kappa}\right)\,\sigma$
and $\left(u,v\right)=\sqrt{1-\sqrt{\kappa}}\,$$\left(x,y\right)$
yields the normalized equations 
\begin{eqnarray}
\hspace{-0.75cm}\partial_{t}A & = & (d+i)\left(\partial_{u}^{2}+\partial_{v}^{2}\right)A+f\left(\left|A\right|^{2}\right)A,\label{eq:Rosa1-1}\\
\hspace{-0.75cm}f\left(P\right) & = & \left(1-i\alpha\right)g\left(1+q\right)h\left(P\right)-\left(1-i\beta\right)qh\left(sP\right)-1,\label{eq:Rosa2-1}
\end{eqnarray}
where we defined $F=f/\left(1-\sqrt{\kappa}\right)$ and $P=|A|^{2}$.

\subsection{Behavior of the homogeneous solution}\label{subsec:Sup_CW}

The monochromatic solutions of Eqs.(\ref{eq:Rosa1}-\ref{eq:Rosa2})
denoted as $A=\sqrt{P}\exp\left(-i\omega t\right)$ with $P\in\mathbb{R}$
are given by $\Re\left(F\right)=0$ which yields the implicit relation
between gain and power 
\begin{eqnarray}
g & = & \frac{1+qh\left(sP\right)}{\left(1+q\right)h\left(P\right)}.\label{eq:SS1}
\end{eqnarray}
As we consider the transverse profile of a LB, $P$ actually corresponds
to the power carried by the temporal LS and the multiple solutions
of Eq.~(17) are to be identified with temporal LSs with
different power densities. Once $P$ is known, the carrier frequency
of the solution can be found independently solving 
\begin{eqnarray}
\omega & = & \alpha g\left(1+q\right)h\left(P\right)-\beta qh\left(sP\right).\label{eq:SS2}
\end{eqnarray}

\paragraph*{Super-subcritical transition point}

A simple Taylor expansion of Eq.~(17) around the threshold
gives the relation 
\begin{eqnarray}
g & = & 1+p\frac{1+q\left(1-s\right)}{2\left(1+q\right)}.
\end{eqnarray}
As such, the solution curve experiences a transition from super-critical
toward sub-critical when 
\begin{eqnarray}
1+q_{c}\left(1-s_{c}\right) & = & 0.
\end{eqnarray}
yielding a relation between the critical saturation $s_{c}$ and the
breadth of the absorber modulation $q_{c}$.

\begin{figure}
\begin{tabular}{cc}
(a)  & (b)\tabularnewline
\includegraphics[width=0.45\columnwidth]{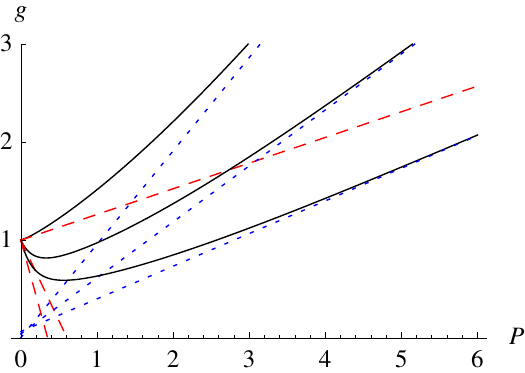}  & \includegraphics[width=0.45\columnwidth]{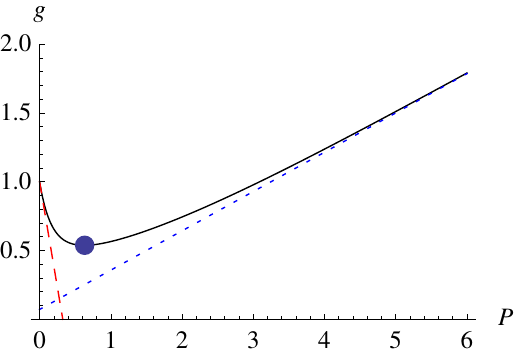}\tabularnewline
\end{tabular}\caption{(a) Evolution of the bistability regime for $s=10$ and increasing
values of $q=0.05,\,0.75$ and $2$ (black lines) and (b) Approximation
of the folding point $\left(P_{m},\,g_{m}\right)$ for $s=10$ and
$q=2.5$ (blue circle). In both cases the homogeneous solution is
represented by a black line and the asymptotic expansions for low
and high power are represented in dashed red and dotted blue lines,
respectively. \label{fig:supersubCW}}
\end{figure}

\begin{figure}
\begin{tabular}{cc}
(a)  & (b)\tabularnewline
\includegraphics[width=0.45\columnwidth]{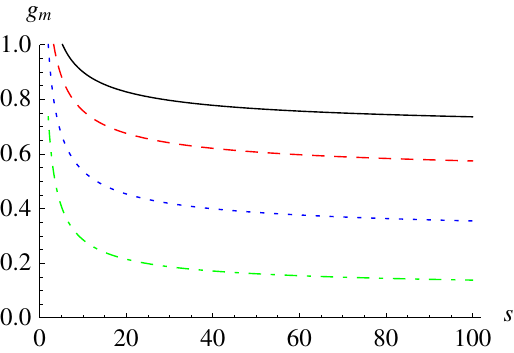}  & \includegraphics[clip,width=0.45\columnwidth]{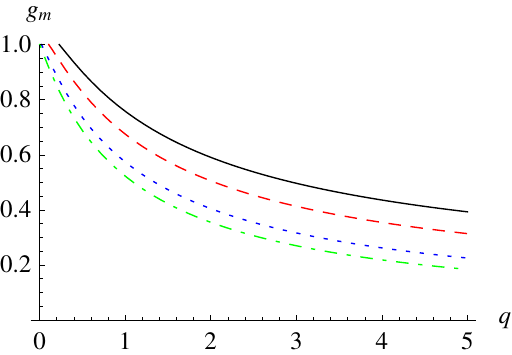} \tabularnewline
\end{tabular}\caption{(a) Evolution of the approximation of the folding point as a function
of $s$ and different values of $q$ that correspond to $q=0.5$ (continuous)
$q=1$ (dashed red), $q=2.5$ (dotted blue) and $q=10$ (dash-dotted
green) lines, respectively. (b) Evolution of the approximation of
the folding point as a function of $s$ and different values of $q$
that correspond to $q=0.5$ (continuous) $q=1$ (dashed red), $q=2.5$
(dotted blue) and $q=10$ (dash-dotted green) lines, respectively.\label{fig:scaling_CW}}
\end{figure}

\paragraph*{High power branch}

The upper solution branch for which the values of $P$ are large,
can be approximated by replacing the nonlinear function $h\left(P\right)$
by its asymptotic value $h\sim1/P$, such that Eq.~(17)
takes the form 
\begin{eqnarray}
g\left(P\right) & \simeq & \frac{P+\frac{q}{s}}{1+q}\sim\frac{P}{1+q}
\end{eqnarray}
and where we used the large saturation approximation. The evolution
of the solution from the super-critical toward the sub-critical regime
is depicted in Fig.~\ref{fig:supersubCW}(a), in addition to the
asymptotic regimes for low and high intensities. 

\paragraph*{Folding point approximation}

The folding point is achieved at powers at which the saturable absorber
is saturated but the gain is not. We replace only $h\left(sP\right)$
by its high power expansion in Eq.~(\ref{eq:SS1}) to find 
\begin{eqnarray}
g\left(P\right) & = & \frac{P+\frac{q}{s}}{\left(1+q\right)\left(1-e^{-P}\right)}.
\end{eqnarray}
Searching for the folding point as a minimum of $g\left(P\right)$
yields a solution $P_{m}$ that is only a function of $q/s$ and reads
\begin{eqnarray}
P_{m}\left(\frac{q}{s}\right) & = & -1-\frac{q}{s}-W_{-1}\left(-e^{-1-\frac{q}{s}}\right)
\end{eqnarray}
with $W_{n}\left(z\right)$ the Lambert-W function. We define the
value of the gain at the folding point as $g_{m}=g\left(P_{m}\right)$
as it is a measure of the extent of the sub-critical region 
\begin{eqnarray}
g_{m} & = & -\frac{W_{-1}\left(-e^{-1-\frac{q}{s}}\right)}{1+q}.\label{eq:gmLambert-1}
\end{eqnarray}

The accuracy of our approximation is depicted in Fig.~\ref{fig:supersubCW}b)
by a blue circle. One can see that is is indistinguishable from the
exact value.

Finally, while the solution of Eq.~(\ref{eq:gmLambert-1}) using
the Lambert function is maybe complicated, the asymptotic values of
$g_{m}$ in the limit of large saturation and large absorption are
simply 
\begin{equation}
\lim_{s\rightarrow\infty}g_{m}=\frac{1}{1+q}\quad,\quad\lim_{q\rightarrow\infty}g_{m}=\frac{1}{s}.\label{eq:scaling_CW}
\end{equation}
The scaling behavior of the folding point as a function $s$ using
Eq.~\ref{eq:gmLambert-1} are represented in Fig.~\ref{fig:scaling_CW}(a).
We note that the asymptotic behavior in Fig.~\ref{fig:scaling_CW}(a)
can only be obtained for very large values of $s$. Similarly, the
curves in Fig.~\ref{fig:scaling_CW}(b) converge toward $g_{m}=s^{-1}$
for unrealistically large values of $q$. However, the behavior predicted
by Eq.~(\ref{eq:scaling_CW}) is qualitatively verified.

\subsection{Bifurcation analysis of the Rosanov Equation }\label{subsec:bif-ana}

\subsubsection{Numerical method}

The LS solutions of Eqs.(\ref{eq:Rosa1}-\ref{eq:Rosa2}) can
be found in the form $A(r,t)=a(r)\,e^{-i\omega t},$where $r=(u,v)$
are normalized transverse spatial coordinates, $a(r)$ is the complex
amplitude with the field intensity $P=|a|^{2}$ localized around some
point in space and $\omega$ is the spectral parameter. To directly
track LS solutions of Eq.~(9) in the parameter space,
we make use of pde2path~\cite{dohnal2014pde2path,pde2path}, a numerical
pseudo-arc-length bifurcation and continuation package for systems
of elliptic partial differential equations over bounded multidimensional
domains which is based on the finite-element methods of Matlab's pdetoolbox
and OOPDE toolbox.

In general, path continuation procedures determine stationary solutions
of a dynamical system combining prediction steps where a known steady
state solution is advanced in parameter space via a tangent predictor
and correction steps where Newton procedures are used to converge
to the next solution at a new value of the primary continuation parameter~\cite{Kuznetsov_Book,Seydel_Book}.
In this way one can start at, e.g., a numerically given solution,
continue it in parameter space, and obtain a solution branch including
its stability. The primary continuation parameter is in our case the
gain parameter $g$, whereas the corresponding spectral parameter
$\omega$ is used as an additional free parameter that is automatically
adapted to the corresponding $g$ during continuation. Further, one
needs an additional auxiliary condition to break the phase shift symmetry
of the system in question in order to prevent the continuation algorithm
to trivially follow solutions along the corresponding degree of freedom.
This condition can be easily implemented by, e.g., setting the phase
of the LS to zero in the center of the computational domain.

To increase computational efficiency in two dimensions, we exploit
the rotational symmetry of the LS and only compute one quarter of
the physical domain $\Omega_{2}=[-L_{u},\,L_{u}]\times[-L_{v},\,L_{v}]$
with $L_{u}=L_{v}=90$ using a grid with $N_{u}\times N_{v}$ mesh
points, $N_{u}=N_{v}=256$. As a result Neumann boundary conditions
are imposed in both $u,v$ directions, whereas periodic boundary conditions
are employed in the one-dimensional case. There, the continuation
is performed on the one-dimensional domain $\Omega_{1}=[-L_{u},L_{u}]$
with $L_{u}=90$ using $N_{u}=512$ equidistant mesh points.

\subsubsection{Evolution of the folding point of the LS with $q$ and $s$}

\begin{figure}
\begin{tabular}{cc}
(a)  & (b)\tabularnewline
\includegraphics[width=0.45\columnwidth]{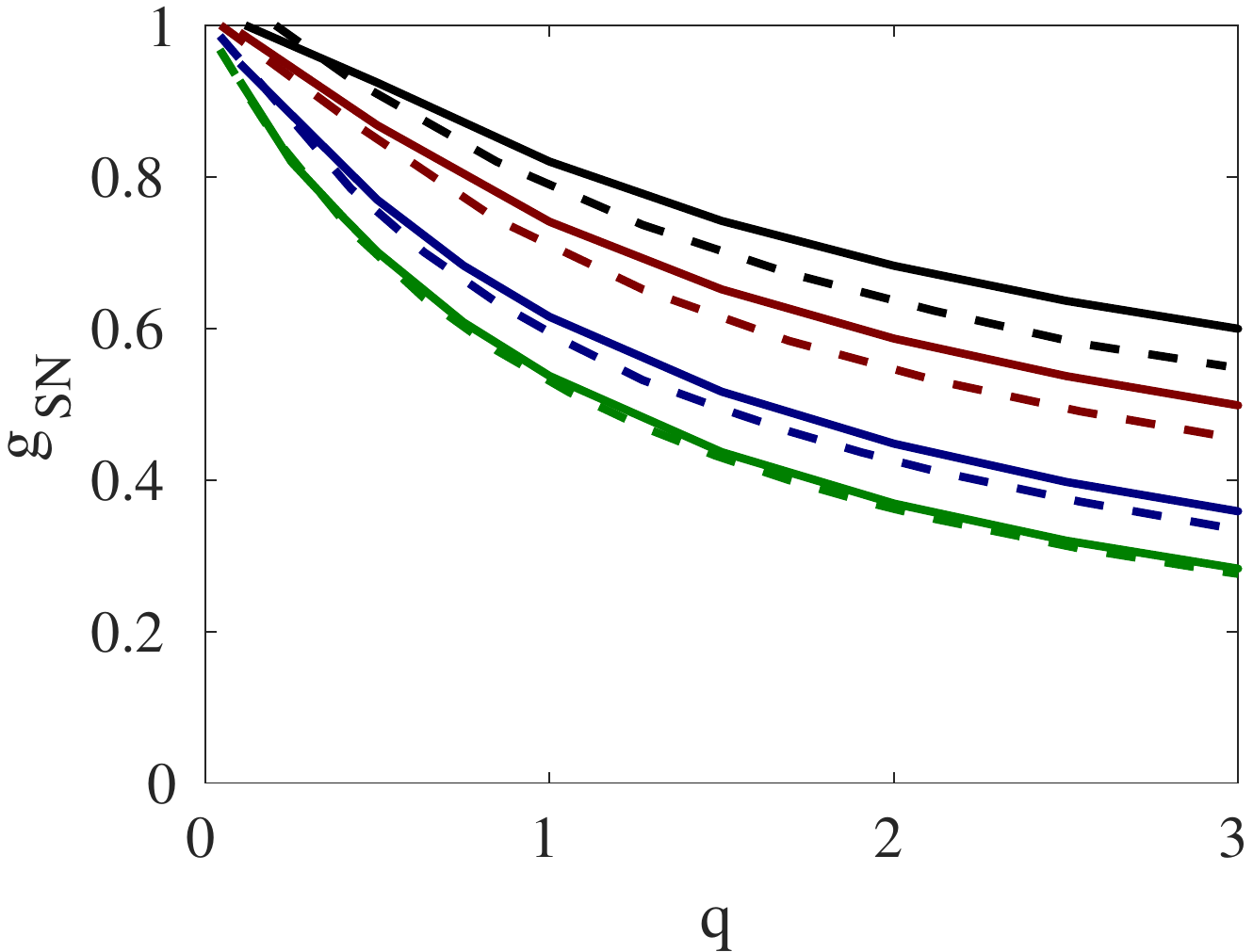}  & \includegraphics[width=0.45\columnwidth]{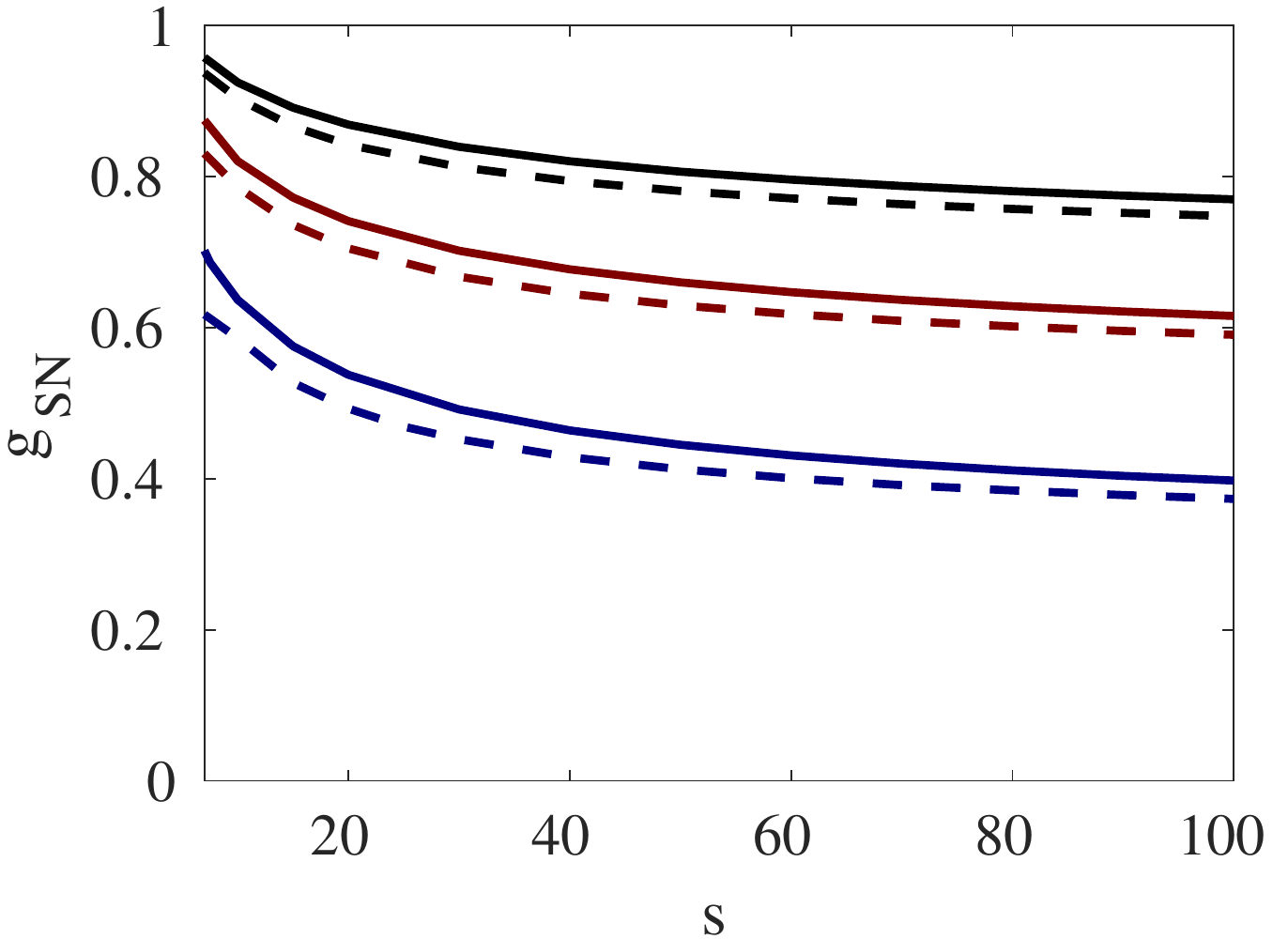}\tabularnewline
\end{tabular}\caption{(Color online) (a) Continuation of the folding point $g_{SN}$ of
a single LS branch as a function of the normalized absorption $q$
and different values of $s$ that correspond to $s=10$ (black) $s=20$
(red), $s=100$ (blue) and $s=1000$ (green). (b) Continuation of
$g_{SN}$ of a single LS branch as a function of $s$ and different
values of $q$ that correspond to $q=0.5$ (black) $q=1$ (red), $q=2.5$
(blue) lines. Dashed (solid) lines in both panels indicate one- (two-)
dimensional continuation, respectively. }
\label{fig:folding_p2p} 
\end{figure}

In this section we discuss the influence of the system parameters,
e.g., the normalized absorption $q$ and the saturation parameter
$s$ on the behavior of the SN point $g=g_{SN}$ of the single LS
branch in both one and two spatial dimensions. To this aim we perform
one- and two-dimensional fold continuations for different $q$ and
$s$ and our results are presented in Fig.~\ref{fig:folding_p2p}.

Figure~\ref{fig:folding_p2p}~(a) shows continuation of the folding
point $g_{SN}$ as a function of $q$, obtained for different values
of $s$, while in panel (b) fold continuation of $g_{SN}$ as a function
of $s$, for different values of $q$, are presented. Dashed (solid)
lines in both panels indicate one- (two-) dimensional continuation,
respectively. Note that the two-dimensional folding point is always
shifted to higher current values compared with the one-dimensional
case, the overall evolution of the folding point $g_{SN}$ remains
the same. Furthermore, the behavior of $g_{SN}$ with $q$ and $s$
follows the same trends as the homogeneous solution (cf. Sec.~\ref{subsec:Sup_CW}),
including its asymptotic behavior in the limit of large saturation
and large absorption. That is, our predictions for the folding point
evolution of the homogeneous solution hold for the folding point of
the LS solution.

\subsubsection{Unfolding of the spiral}

\begin{figure}
\includegraphics[width=1\columnwidth]{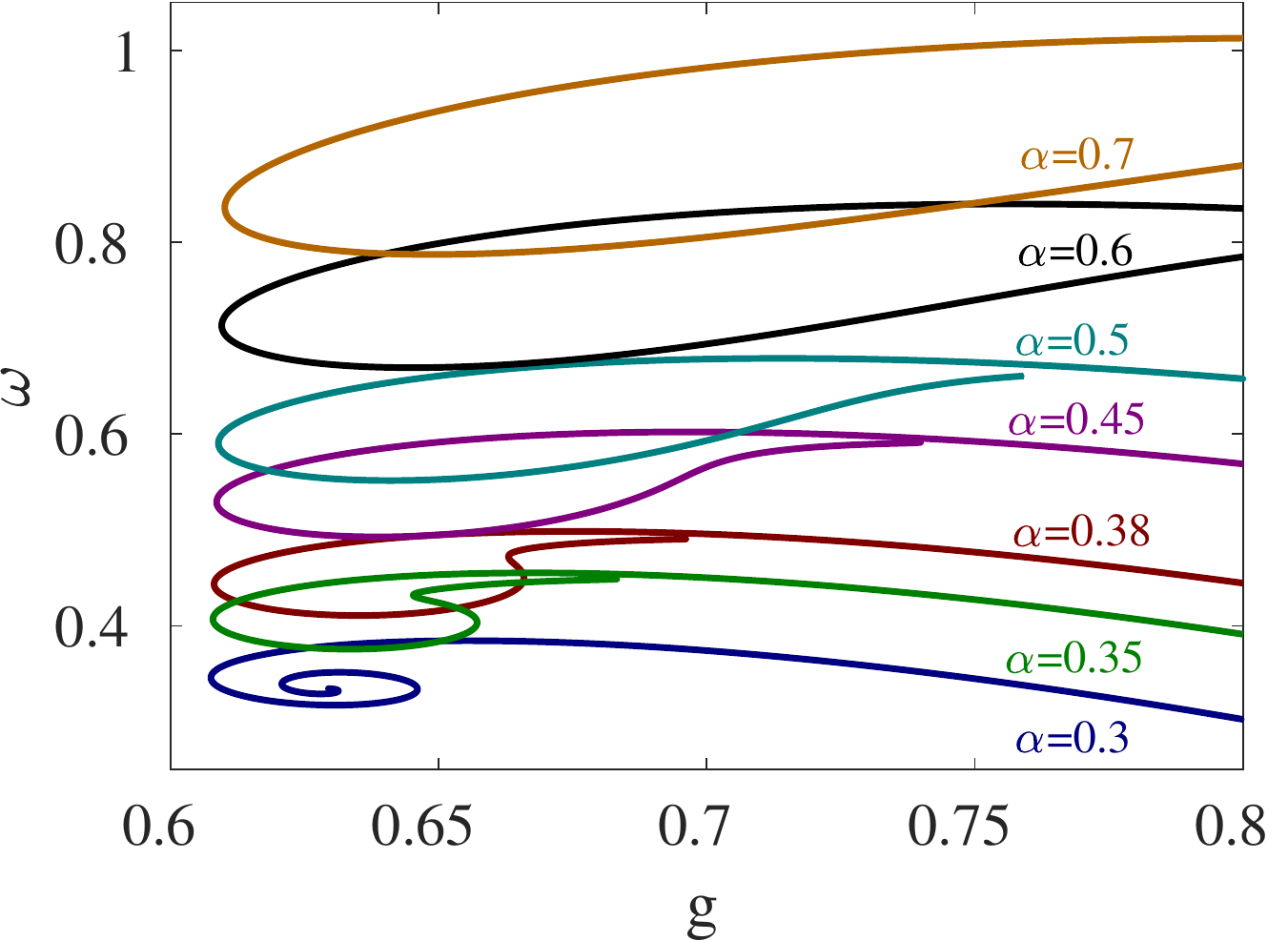}\caption{(Color online) One-dimensional bifurcation diagram of Eq.~9
in $(g,\omega)$ plane calculated for increasing values of $\alpha$
at fixed $\beta=0.5$. For small $\alpha$ the branch of a single
LS forms a spiral which unfolds for increasing $\alpha$. Different
colors correspond to different values of $\alpha$. Other parameters
are the same as in Fig.~\ref{fig:LS1d_vs_CW}. \label{fig:unfolding_spiral}}
\end{figure}

Figure~\ref{fig:unfolding_spiral} shows the typical spiral shape
obtained for the branch of a LS in the $(g,\omega)$ plane at low
values of $\alpha$. One can see that the spiral shape of the LS branch
remains well preserved for small values of $\alpha$ (cf. Fig.~\ref{fig:unfolding_spiral}
for $\alpha=0.3$). However, for increasing $\alpha$ the morphology
of the branch changes: Beyond a certain threshold, the end point of
the spiral curve moves to higher gain values (see Fig.~\ref{fig:unfolding_spiral}
for $\alpha=0.35\ldots0.5$), leading to the unfolding of the spiral
($\alpha=0.6,0.7$). There, the resulting branch bifurcates from the
threshold $g=1$, exhibits a fold at a certain $g$ and continues
towards higher gain values, i.e., it coexists with the upper sub-branch
and always exhibits smaller $\omega$. This branch morphology remains
for ascending $\alpha$ as shown in Fig.~\ref{fig:LS1d_vs_CW}.

\subsubsection{Numerical simulation of the Haus Equation} \label{subsec:Num-sim-Haus}

We solved Eqs.~(\ref{eq:VTJ1}-\ref{eq:VTJ3}) by adding two additional free parameters $\omega$ and $\upsilon$ which correspond
to the the oscillation frequency and the drift velocity of the solution
along the propagation axis. In other words, the round-trip of the
LBs presents a small deviation with respect to the cold cavity round-trip
time that corresponds to a drift $\upsilon$ in the reference frame
of the cold cavity where one takes a snapshot every round-trip time,
see \cite{JCM-PRL-16} for more details. As such Eq.~1 is rewritten as 
\begin{eqnarray}
\partial_{\sigma}E & = & \left(O_{1}+O_{2}\right)E\label{eq:EOp}
\end{eqnarray}
with the two operators $O_{1,2}$ defined as 
\begin{eqnarray}
O_{1} & = & \sqrt{\kappa}\left[\frac{1-i\alpha}{2}G\left(r_{\perp},z,\sigma\right)-\frac{1-i\beta}{2}Q\left(r_{\perp},z,\sigma\right)\right]\nonumber \\
 & + & \sqrt{\kappa}-1+i\omega\\
O_{2} & = & \frac{1}{2\gamma^{2}}\partial_{z}^{2}+\left(d+i\right)\Delta_{\perp}+\upsilon\partial_{z}
\end{eqnarray}

The values of the free parameters $\left(\omega,\upsilon\right)$
are adapted during the time integration via a simple control loop
allowing to determine the frequency by looking at the phase variation
of the peak of the LB and the drift along the propagation axis of
the intensity profile averaged over the transverse dimension(s). In
particular, canceling the natural drift of the solution along the
propagation axis by a proper value of $\upsilon$ allows to maintain
the position of the LB close to the center of the numerical domain.

We solved Eq.~\ref{eq:EOp} using a semi-implicit split-step method
in which the spatial operator $O_{2}$ is evaluated in Fourier space.
Denoting $\tilde{O}_{2}$ the differential operator in Fourier space,
the update sequence reads 
\begin{eqnarray}
E_{1} & = & E_{n}\left(1+\Delta t\times O_{1}\right),\\
E_{2} & = & \mathcal{F}^{-1}\left[e^{\Delta t\times\tilde{O}_{2}}\mathcal{F}\left[E_{1}\right]\right],\\
E_{n+1} & = & E_{2}/\left(1-\Delta t\times O_{1}\right).
\end{eqnarray}
As the carrier variables $G$ and $Q$ are not depending on the slow
time, since we exploited the long cavity limit, the operator $O_{1}$
can be obtained from the integration of Eqs.~(\ref{eq:VTJ2}-\ref{eq:VTJ3})
knowing the pre-existing field intensity distribution and using Dirichlet
boundary conditions at the beginning of the integration domain that
reads
\begin{eqnarray}
\left(G,Q\right)\left(r_{\perp},z=0\right) & = & \left(G_{0},Q_{0}\right).
\end{eqnarray}
We used a physical domain $\Omega_{H}=[-L_{x},\,L_{x}]\times[-L_{y},\,L_{y}]\times[-L_{z},\,L_{z}]$
with $L_{x,y}=160$ and $L_{z}=3$ using a grid with $N_{x}\times N_{y}\times N_{z}$
mesh points and $N_{x,y,z}=128$. Periodic boundary conditions are
automatically imposed in both $x,y$ directions, as a consequence
of the differentiation in Fourier space. We applied standard dealiasing
with a 2/3 rule to the Fourier operator. As the Fourier operator $\mathcal{O}_{2}$
is linear, no dealiasing is needed. As such, one has to put dealiasing
into $\mathcal{O}_{1}$ every time one goes into the Fourier space
and we apply the 2/3 rule in $\mathcal{F}\left[E_{1}\right]$ in Eq.
30 only.

The time step was $\delta t=0.1$ which gives indistinguishable results
for the LB parameters like energy, width, frequency and velocity as
compared to smaller time steps. The main reason for this strong convergence
property stems from the fact that around a steady state $O_{1}\simeq0$
while the spatial operator $O_{2}$ is obtained via exponential differencing
and is therefore exact. As such, the remaining errors stems only from
the operator splitting. It is proportional to a commutator that reads
$\left[1\pm\Delta t\times O_{1},e^{\Delta t\times\tilde{O}_{2}}\right]\simeq0$
since $O_{1}\simeq0$ at steady state. A convergence analysis of the
numerical method yielded the expected second-order accuracy. Finally,
we added to the field equation white Gaussian noise with variance
$\xi=10^{-4}$ mainly to accelerate the escape from unstable solutions
and to avoid the detection of false positive. The time integration
was $\Delta\sigma=3\times10^{3}$. 

The numerical bifurcation diagrams were obtained by increasing $g$
from a minimal value $g=0.4$ up to $g=1$ with a step of $\delta g=2\times10^{-3}$.
As initial conditions, we used a spherical solution $E_{i}\left(x,y,z\right)=E_{0}$
with $\left(x,y,z\right)$ verifying the relation
\begin{equation}
\left(\frac{x}{L_{x}}\right)^{2}+\left(\frac{y}{L_{y}}\right)^{2}+\left(\frac{z}{L_{x}}\right)^{2}<1
\end{equation}
with parameters $L_{x}=L_{y}=10$, $L_{z}=0.26$ and $E_{0}=1$. After
an integration time of $\Delta\sigma$, if a stable LB solution is
found, it is used as an initial condition for the next value of $g$.
After the entire upward scan in $g$ the solution branch is further
extended via continuation downward from the first point that was found
using the spherical IC. This methods allows to get a good approximation
of the folding point limiting the branch at low values of $g$. However,
during such ``blind'' parameters sweeps, a large fraction of the
simulations consist in a dynamics in which the field goes down to
$E=0$ or the spatio-temporal profile explodes and invades the full
numerical domain. As such, special flags were introduced to cut the
time integration. This simple procedure diminishes the computation
times by several orders of magnitudes. The scans along $q$ were performed
with a step $\delta q=3\times10^{-2}$. Simulations were performed
using 100 cores of Xeon E5 CPUs on a HPC cluster Bull B510.

\begin{figure*}
\includegraphics[bb=1000bp 50bp 9400bp 2700bp,clip,width=2\columnwidth]{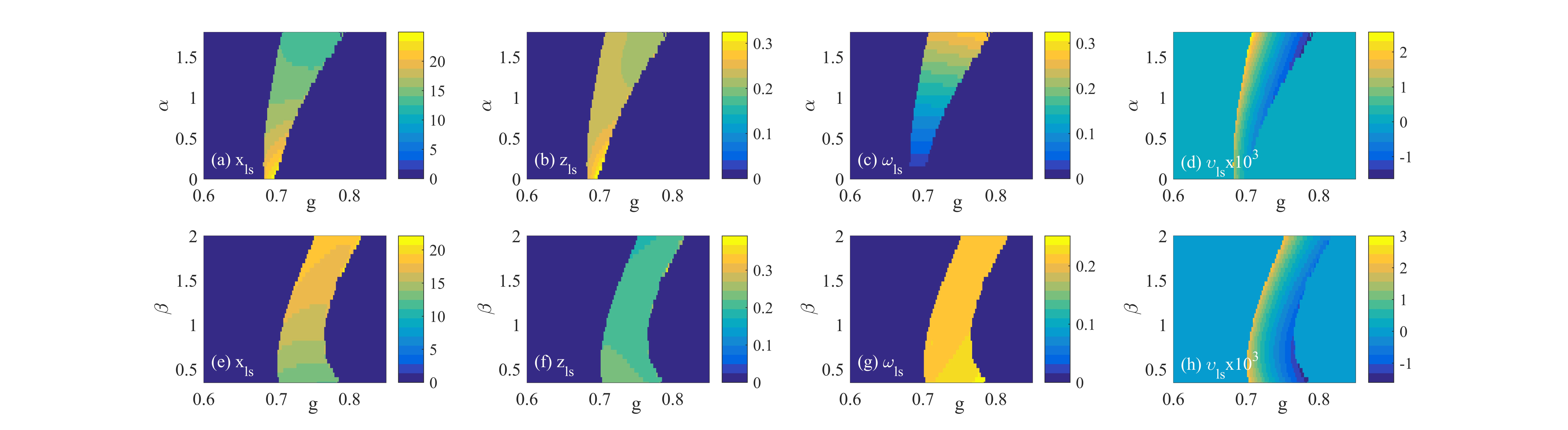}\caption{(Color online). Two-dimensional bifurcation diagrams showing the region
of stable existence of the LBs by integrating the two-dimensional
Haus equation Eqs.~(\ref{eq:VTJ1}-\ref{eq:VTJ3}) as a function of the
gain $g$ and the parameters $\alpha$ (a) and $\beta$ (b). We represent
the FWHM in $x$ (a, e), the FWHM in $z$ (b, f) the frequency of
the solution (c, g) and the drift velocity (d, h). As mentioned in
the main text, the range of stability increases with $\alpha$ and
reduces for small values of $\beta$ in agreement with Fig.~\ref{fig:bif_diag_2d}.
Parameters are $\left(\gamma,\kappa,\Gamma,Q_{0}d\right)=\left(40,0.8,0.04,0.3,10^{-2}\right)$
and $\beta=0.5$ (a-d), $\alpha=1.5$ (e-h).\label{fig:bif_diag_Haus_2d_gab_extra}}
\end{figure*}

The results of the scan in the $\left(\alpha,g\right)$ and $\left(\beta,g\right)$
planes are presented in Fig.~\ref{fig:bif_diag_Haus_2d_gab_extra}
where we represented the characteristics of the LBs, as for instance
their FWHM in the transverse and longitudinal directions, the residual
frequency and the velocity of propagation along the propagation axis.
As mentioned in the main text, the stability region increases with
larger values of $\alpha$ and smaller values of $\beta$.

\begin{figure*}
\includegraphics[bb=1000bp 50bp 10500bp 4500bp,clip,width=2\columnwidth]{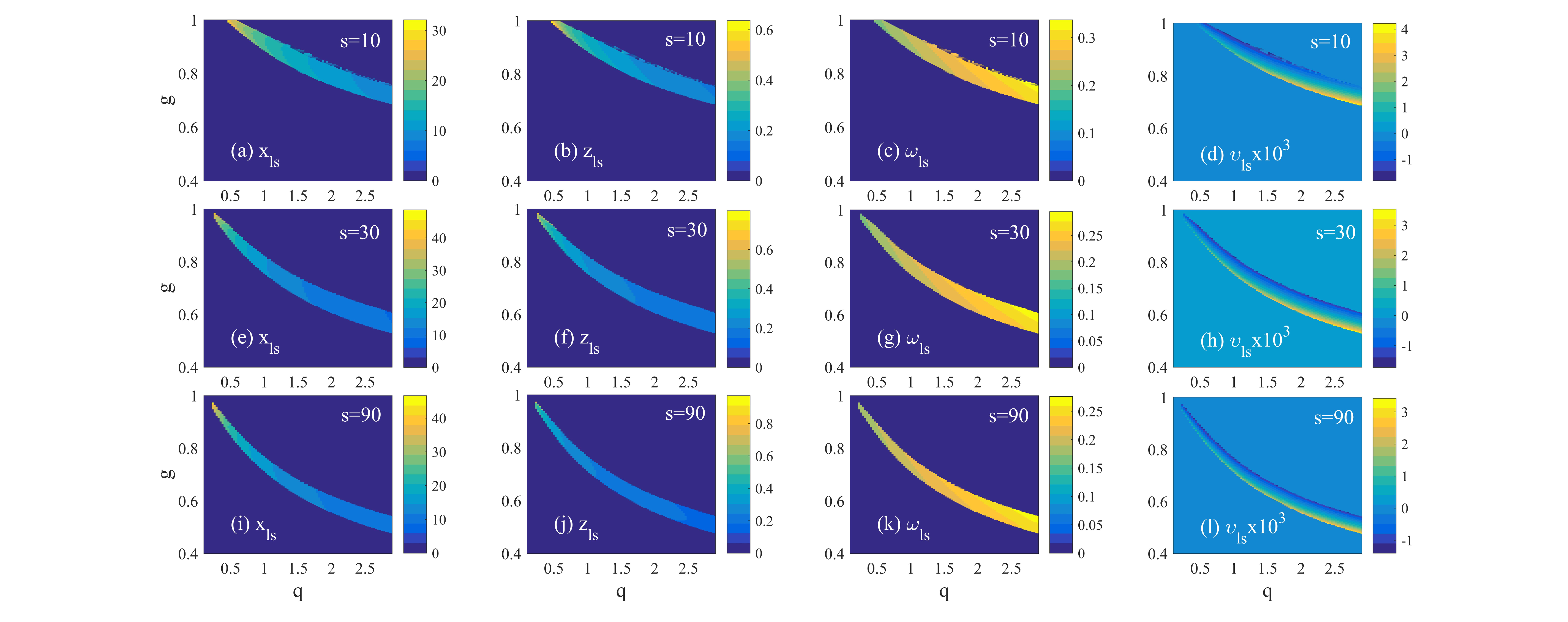}\caption{(Color online). Two-dimensional bifurcation diagram showing the region
of stable existence of the LBs by integrating the two-dimensional
Haus equation Eqs.~(\ref{eq:VTJ1}-\ref{eq:VTJ3}) as a function of the
gain $g$ and the parameters $\alpha$ (a) and $\beta$ (b). We represent
the FWHM in $x$ (a, e), the FWHM in $z$ (b, f) the frequency of
the solution (c, g) and the drift velocity (d, h). As mentioned in
the main text, the range of stability increases with $\alpha$ and
reduces for small values of $\beta$ in agreement with Fig.~\ref{fig:bif_diag_2d}.
Parameters are $\left(\gamma,\kappa,\Gamma,Q_{0}d\right)=\left(40,0.8,0.04,0.3,10^{-2}\right)$
and $\beta=0.5$ (a-d), $\alpha=1.5$ (e-h).\label{fig:bif_diag_Haus_2d_sgq_extra}}
\end{figure*}

\end{document}